\pdfoutput=1
\documentclass[fleqn,usenatbib]{mnras}

\usepackage{newtxtext,newtxmath}
\usepackage[T1]{fontenc}
\usepackage{ae,aecompl}

\usepackage{graphicx}	% Including figure files
\usepackage{amsmath}	% Advanced maths commands
\usepackage{amssymb}	% Extra maths symbols
\usepackage{float}
\usepackage{gensymb}
\usepackage{tabu}
\usepackage[final]{animate}
\usepackage{enumerate}
\title[The fossil record of MaNGA spirals]{\textit{SDSS-IV MaNGA}: Excavating the fossil record of stellar populations in spiral galaxies}

\author[T. Peterken et al.]{
Thomas Peterken,$^{1}$\thanks{E-mail: Thomas.Peterken@nottingham.ac.uk}
Michael Merrifield,$^{1}$
Alfonso Arag{\'o}n-Salamanca,$^{1}$
\newauthor
Amelia Fraser-McKelvie,$^{1}$ Vladimir Avila-Reese,$^{2}$ Rog{\'e}rio Riffel,$^{3,4}$ Johan Knapen,$^{5,6}$
\newauthor
Niv Drory$^{7}$
\\
$^{1}$School of Physics and Astronomy, University of Nottingham, University Park, Nottingham NG7 2RD, UK\\
$^{2}$Instituto de Astronom{\'i}a, Universidad Nacional Aut{\'o}noma de M{\'e}xico, A.P. 70--264, 04510 CDMX, M{\'e}xico\\
$^{3}$Instituto de F{\'i}sica (IF), Universidade Federal do Rio
Grande do Sul, CP 15051, Porto Alegre 91501-970, RS, Brazil \\
$^{4}$Laborat\'orio Interinstitucional de e-Astronomia - LIneA, Rua
Gal. Jos\'e Cristino 77, Rio de Janeiro, RJ - 20921-400, Brazil\\
$^{5}$Instituto de Astrof\'{i}sica de Canarias, V\'{i}a L\'{a}ctea S/N, E-38205 La Laguna, Spain\\
$^{6}$Departamento de Astrof\'{i}sica, Universidad de La Laguna, E-38206 La Laguna, Spain \\
$^{7}$McDonald Observatory, The University of Texas at Austin, 1 University Station, Austin, TX 78712, USA
}

\date{Draft copy}

\pubyear{2019}

\begin{document}
\label{firstpage}
\pagerange{\pageref{firstpage}--\pageref{lastpage}}
\maketitle

% Abstract of the paper
\begin{abstract}
We perform a ``fossil record'' analysis for $\approx 800$ low-redshift spiral galaxies, using \textsc{Starlight} applied to integral field spectroscopic observations from the SDSS-IV MaNGA survey to obtain fully spatially-resolved high-resolution star formation histories (SFHs).  From the SFHs, we are able to build maps indicating the present-day distribution of stellar populations of different ages in each galaxy.  We find small negative mean age gradients in most spiral galaxies, especially at high stellar mass, which reflects the formation times of stellar populations at different galactocentric radii.  We show that the youngest ($<10^{8.5}$~years) populations exhibit significantly more extended distributions than the oldest ($>10^{9.5}$~years), again with a strong dependence on stellar mass.
%Since radial migration effects are likely to be small, these results are consistent with inside-out growth being a generic feature of spiral galaxies, and is most significant in massive galaxies.
By interpreting the radial profiles of ``time slices'' as indicative of the size of the galaxy at the time those populations had formed, we are able to trace the simultaneous growth in mass and size of the spiral galaxies over the last 10~Gyr.  Despite finding that the evolution of the measured light-weighted radius is consistent with inside-out growth in the majority of spiral galaxies, the evolution of an equivalent mass-weighted radius has changed little over the same time period.  Since radial migration effects are likely to be small, we conclude that the growth of disks in spiral galaxies has occurred predominantly through an inside-out mode (with the effect greatest in high-mass galaxies), but this has not had anywhere near as much impact on the distribution of mass within spiral galaxies.
\end{abstract}

% Select between one and six entries from the list of approved keywords.
% Don't make up new ones.

%%%%%%%%%%%%%%%%%%%%%%%%%%%%%%%%%%%%%%%%%%%%%%%%%%

%%%%%%%%%%%%%%%%% BODY OF PAPER %%%%%%%%%%%%%%%%%%

\section{Introduction}
\label{sec:Intro}

Understanding how, when, and where galaxies built their mass is key to cosmology and astronomy.  Analysis of the evolution of the masses and sizes of galaxies has generally been limited to comparisons of different galaxy populations at different redshifts.  Studies done in this manner have shown that galaxies have grown in radius whilst building their mass (e.g.\ \citealt{Trujillo+07, vanderWel+08, vanDokkum+08, vanDokkum+13, Patel+13, Papovich+15, Whitney+19}), giving rise to the concept of ``inside out'' formation.  It is thought that such growth in the most massive galaxies has been due to some combination of multiple minor mergers \citep{Naab+09, Furlong+17}, gas accretion \citep{Conselice+13}, and quasar feedback \citep{Fan+08}.  These approaches have given us a good insight into how the average properties of galaxies have evolved over cosmic time, but because we cannot track the evolution of any individual system in this way, it is difficult to go beyond such global properties.  Although some studies of galaxies at different redshifts have managed to show inside-out growth in disk-like galaxies (e.g.\ \citealt{Trujillo+06, Patel+13}), most are restricted to the highest mass galaxies, so this picture of inside-out growth is normally limited to early-type galaxies.

An alternative approach which is more suited to late-type galaxies is to explore the stellar populations in different regions of a galaxy, particularly through studying how the mean stellar age varies with radius.  This method requires high-quality spectral data at multiple locations across the face of a galaxy, and so has only been undertaken in detail for large numbers of galaxies since the advent of integral-field spectroscopic surveys such as the Calar-Alto Legacy Integral Field Array (CALIFA; \citealt{CALIFA}), Sydney-AAO Multi-object Integral field spectrograph (SAMI; \citealt{SAMI}), and Mapping Nearby Galaxies at APO (MaNGA; \citealt{Bundy+15}) surveys.  Using such a ``fossil record'' approach applied to integral-field spectroscopic data has revealed that most galaxies exhibit negative age gradients (e.g. \citealt{Mehlert+03, SanchezBlazquez+14, GonzalezDelgado+15, Goddard+17}) --- with younger outskirts than centres --- or earlier formation times of the central regions (e.g. \citealt{IbarraMedel+16}) providing more evidence for a dominant ``inside out'' growth mode occurring in galaxies of all Hubble types.  This is also backed up by \citet{Sacchi+19} for the case of NGC~7793, who find that broad-band observations of resolved stellar populations in this nearby spiral galaxy indicate a clear gradient in stellar age.

By applying stellar population modelling methods to integral-field spectroscopic data from the CALIFA survey, \citet{CidFernandes+13, CidFernandes+14}, \citet{Perez+13}, \citet{GonzalezDelgado+17}, and \citet{GarciaBenito+19} have shown that it is possible to reveal much more about a galaxy's history by deriving full star-formation histories rather than mean ages.  We have shown previously that such analyses of the spatial variation in stellar populations of spiral galaxies can help us understand the structure of the spiral arms and bars, but here we investigate how such approaches can also help us study the evolution and growth of populations of galaxies.

Comparative studies of the masses and sizes of galaxies at different lookback times are most effective to measure the growth of early-type galaxies since these are typically the most massive and luminous objects at any given redshift so are easy to identify.  By contrast, a fossil record analysis acts as a complementary approach best suited to --- but by no means limited to; see e.g.\ \citet{Lacerna+20} --- studying the growth of late-type galaxies, as such galaxies have in general had continued growth over the last several~Gyr.  This extended star-formation in late-type galaxies can be traced using fossil record methods, providing that care is taken to ensure that older populations can be detected when the flux may be dominated by the younger and brighter populations.  Of course, there exists a population of spiral galaxies which are passive (see for example \citealt{Masters+10, FraserMcKelvie+16}) --- contrary to the well-known relation between the morphology and star formation rate \citep{TullyMouldAaronson82, Baldry+04} --- so a morphological classification does not always define the extent of the star-formation history of each galaxy.  However, for consistency, we have chosen to study a galaxy population selected on their morphology rather than colour, to better understand how this well-defined galaxy class have evolved over time.

Here, we perform full spectral fitting of spiral galaxies from the MaNGA survey \citep{Bundy+15} and measure spatially-resolved star formation histories, to uncover their formation sequences.

\bigskip

This paper is structured as follows.  In \S\ref{sec:Data}, we outline the data we use from the MaNGA survey.  In \S\ref{sec:SampleSelection} we describe how a sample of spiral galaxies from the MaNGA target list was selected, and in \S\ref{sec:Fitting} we detail the spectral fitting method employed (with some tests of this method outlined in Appendix~\ref{app:FittingTests}).  We then describe how the derived star-formation histories are processed in \S\ref{sec:AgeSlicing}.  The mean age and metallicity gradients are derived in \S\ref{sec:MeanAgeMets}.  In \S\ref{sec:MassBuildupTime} and \S\ref{sec:Concentrations} we analyse the star-formation histories and spatially-resolved stellar populations in more detail, and infer the evolution of the mass-size relation in \S\ref{sec:MassSize}.  Finally, we discuss the interpretation and context of the results in \S\ref{sec:Discussion}.

\section{Data}
\label{sec:Data}

\subsection{MaNGA}
\label{sec:DataMaNGA}

MaNGA \citep{Bundy+15} is part of the fourth generation of the Sloan Digital Sky Survey (SDSS-IV; \citealt{Blanton+17}).  By the survey's completion is 2020, MaNGA will have acquired 2.5~arcsec resolution integral-field spectroscopic observations of more than 10,000 galaxies in the redshift range $0.01<z<0.15$ \citep{Yan+16-design}.  The survey makes use of the BOSS spectrograph \citep{Smee+13} on the 2.5-metre SDSS telescope \citep{Gunn+06} at the Apache Point Observatory, which has a spectral resolution of $R\approx2000$ and covers a large wavelength range of 3600-10300~\AA.  The raw data's calibration is described by \citet{Yan+16-cal}, with the datacubes then reduced using MaNGA's data reduction pipeline (DRP; \citealt{Law+16}).  For each target galaxy, observations are taken out to at least either 1.5~$R_{\textrm{e}}$ or 2.5~$R_{\textrm{e}}$ (to form the ``Primary'' and ``Secondary'' samples respectively; \citealt{Law+15}), where $R_{\textrm{e}}$ is the elliptical half-light radius measured photometrically by the NASA-Sloan Atlas \citep{NSA}.  This is achieved using integral-field units of five different sizes, from the 127-fibre IFUs with a diameter of 32~arcsec, to the 19-fibre IFUs of 12~arcsec diameter \citep{Drory+15}.

The MaNGA target selection was chosen to obtain a flat distribution in log(stellar mass) \citep{Wake+17}.  Neither the Primary nor Secondary samples are therefore volume-limited; instead the high-mass galaxies are over-represented while the low-mass galaxies are under-represented.  The Primary+ (``colour-enhanced'') sample is an extended Primary sample but with an oversampling of the ``green valley'' galaxies \citep{Wake+17} so is therefore also unrepresentative in this way too.  However, since the sample selection in all cases is well-defined \citep{Wake+17}, a weighting has been determined for each galaxy to correct for these selection biases and form a representative volume-limited sample (referred to as the ``Primary+ sample weighting'' throughout this paper).

In this work, we make use of some of the analysis outputs of MaNGA's data analysis pipeline (DAP; \citealt{DAP}).  Specifically, we use the measured stellar velocities $v_{\star}$, deprojected radii $R$, and emission line spectra [which are themselves described in detail by \citet{DAPLines}], all of which are derived using full spectral modelling.  The data we use here is from the internal MaNGA product launch 8 (MPL-8) data release, which contains completed observations of 6778 galaxies.

\subsection{Galaxy Zoo}

We also make use of the morphological classifications of each MaNGA galaxy provided by volunteer ``citizen scientists'' as part of Galaxy Zoo \citep{Lintott+08, Lintott+11}.  The second phase of the project (Galaxy Zoo 2, hereafter GZ2; \citealt{Willett+13}) includes publicly-available detailed classifications of galaxies based on SDSS DR7 imaging.  The users' classifications are weighted and combined to obtain a consensus fraction for each answer to each question for each galaxy, using methods described by \citet{Willett+13} and \citet{Hart+16}.  We use the redshift-debiased and user-weighted probabilities --- which we denote as $p_\textrm{classification}$ --- from the \citet{Hart+16} catalogue.

\section{Sample selection}
\label{sec:SampleSelection}

A sample of spiral galaxies was drawn from the MPL-8 data release using the recommendations of \citet[Table 3]{Willett+13};  see also \citealt{Masters+19} for another recent implementation.  We first remove the 45 galaxies in the matched MPL-8/GZ2 catalogues that more than 50\% of GZ2 users have classified as having some form of star or artifact in the image.  To filter out elliptical galaxy morphologies, we select the 4201 galaxies with $p_\textrm{features or disk} > 0.43$ and at least 20 classifications in this question, as recommended by \citet{Willett+13}.

Since we are interested in the variation in stellar population properties across the face of each spiral galaxy, we remove edge-on galaxies from this sample.  This cut can be made with either the GZ2 classifications --- specifying $p_\textrm{not edge-on} > 0.8$ --- following \citet{Willett+13}, or using an axis ratio cut --- requiring $\frac{b}{a} \geq 0.4$ --- following \citet{Hart+17}.  To select only face-on galaxies, we choose galaxies that satisfy the \citet{Willett+13} criterion and have an axis ratio of $\frac{b}{a} \geq 0.5$ (corresponding to an inclination of $i \leq 60\degree$ assuming the galaxies can be modelled as a thin intrinsically circular disks).  We used this higher axis ratio cut compared to that used by \citet{Hart+17} to ensure that we have selected only galaxies for which the radial structure is clearly resolvable with MaNGA.  Of the 5902 MPL-8 galaxies for which GZ2 classifications are available, this leaves a sample of 1686 close-to-face-on disky galaxies.  Of these, 1314 galaxies satisfy the \citet{Willett+13} requirement for spiral galaxies of $p_\textrm{spiral}>0.8$ and 20 individual classifications in this question.

We then remove 109 galaxies which have flags for bad or questionable-standard data in the MaNGA DRP, or for which the MaNGA MPL-8 DAP dataproducts are unavailable.  To ensure consistency in the spatial resolution relative to the galaxy size, we remove galaxies which are part of MaNGA's Secondary sample.  For the final sample of spiral galaxies, we therefore select only those 795 which are in the Primary+ MaNGA sample, for which MaNGA observations extend to at least 1.5~$R_\textrm{e}$.  The median redshift of galaxies in our sample (weighted by the MaNGA Primary+ sample weighting) is $z=0.026$, and 75\% of the (weighted) sample are at redshifts $z<0.03$.

\section{Spectral fitting}
\label{sec:Fitting}

Using a similar technique to that employed in \citet{Peterken+19PS} and \citet{Peterken+19TS}, we fit each spectrum in each galaxy using \textsc{Starlight} \citep{Starlight}.  We first de-redshift and subtract the emission-line spectrum using the MaNGA DAP \citep{DAP, DAPLines}, and then fit using E-MILES \citep{E-MILES} single stellar population (SSP) templates.  The de-redshifted and emission-subtracted MaNGA spectra are rebinned onto a linear wavelength scale (as required by \textsc{Starlight}) before fitting.  \textsc{Starlight} then uses an iterative method to find the best-fit linear combination of the input templates, and returns the relative weights given to each SSP template in the fit, along with line-of-sight velocity $v_{\star}$, velocity dispersion $\sigma_{\star}$, and the amount of dust reddening $A_{\textrm{V}}$.  

\subsection{Template stellar population spectra}
\label{sec:FittingSSPs}

We use a combination of 9 ages (log(age/years) = 7.85, 8.15, 8.45, 8.75, 9.05, 9.35, 9.65, 9.95, 10.25) and 6 metallicities ($[\textrm{M}/\textrm{H}] = -1.71, -1.31, -0.71, -0.40, +0.00, +0.22$) from the standard E-MILES library (\citealt{E-MILES}, based on the earlier MILES library of \citealt{Vazdekis+10}), assuming a \citet{Chabrier03} IMF, \citet[``Padova'']{Girardi+00} isochrones, and Milky-Way [$\alpha/\textrm{Fe}$] (``baseFe'').  To sample the full star-formation histories, we also include the younger templates of \citet{Asa'd+17} covering 6 ages ($\log(age/\textrm{years}) = 6.8, 6.9, 7.0, 7.2, 7.4, 7.6$) and the two recommended metallicities ($[\textrm{M}/\textrm{H}] = -0.41, +0.00$), which are generated using the same method as the E-MILES set of \citet{E-MILES}, but with the earlier \citet{Bertelli+94} version of the Padova isochrones.  Combining these libraries allows us to exploit the high spectral resolution of both MaNGA and E-MILES templates, while still being able to fully fit the whole of the star-formation histories of star-forming regions without combining different libraries produced in completely different ways.

\subsection{\textsc{Starlight} configuration}
\label{sec:FittingSLparams}

We use \textsc{Starlight} in a ``long fit'' mode to prioritise robustness over computation time, based on the recommendations from extensive testing of \textsc{Starlight} by \citet{Ge+18} and \citet{FakeNews}.  We limit the fit to the wavelength range of 3541.4 to 8950.4~\AA, where the raw E-MILES templates have a constant FWHM of 2.51~\AA.  To ensure that the model and measured spectra have consistent resolution, we degraded each of the SSP templates to the wavelength-dependent resolution of the median spaxel spectrum from all galaxies in our sample, using the line spread function measured by the DAP \citep{DAP, DAPLines}.

Since the DAP robustly models the emission lines \citep{DAPLines}, we use \textsc{Starlight} in its ``NOCLIP'' mode to ensure that all of the diagnostic absorption lines are fully fitted.  To ensure that the star-formation history of each spaxel is measured as fully as possible (defined here as the mass weights assigned to each SSP template divided by the time interval between that template and the next-youngest one), we require \textsc{Starlight} to retain at least 97\% of its fit's total light during the ``EX0'' phase of reducing the number of templates used in the final fit (i.e. EX0s\_method\_option~= CUMUL, EX0s\_Threshold~= 0.03).  This configuration helps to recover the presence of older stellar populations even when their flux has been obscured by the presence of a younger population, for example.  The light weights we use from the fits are those contributed by each template at 4020~\AA\ (as used by \citet{Starlight}), and we allow the sum of weights at this wavelength to be between 50\% and 150\% of the input spectrum\footnote{Any given model may represent the entire spectrum well at all wavelengths \textit{except} that at which we measure the weights.  It is therefore possible for a good fit's light weights at any chosen wavelength not to sum to exactly 100\% of the input spectrum, so allowing this large discrepancy in the weights' sum ensures that the best fit of the whole spectrum is used.}.  In all subsequent analysis, we use either the mass weights (using the implicit mass-to-light ratios included in the E-MILES SSP models and assuming a flat $\Lambda$CDM cosmology with $\textrm{H}_{0} = 70$~km/s) or only compare the spatial variation of flux weights of a specific age, rendering the exact choice of reference wavelength irrelevant.

From the \textsc{Starlight} mass weights, a measure of the total stellar mass within the MaNGA FOV can be readily calculated.  Reassuringly, we find that these stellar masses agree well with those measured by the NASA-Sloan Atlas \citep{NSA}, but with a small offset due likely due to the difference in FOV limitations.  We discuss this comparison further in \S\ref{sec:MassSize}.  Although it is possible that the \citep{NSA} stellar mass measurements are more robust than the \textsc{Starlight}-derived measurements, the consistency between the two measurements is close enough to allow us to use either one.  However, in measuring the mass growth in \S\ref{sec:MassSize}, we are limited to using the \textsc{Starlight} measurements.  Therefore, for consistency, any quoted galaxy stellar mass measurements are those measured by \textsc{Starlight} unless stated otherwise.  The E-MILES library contain stellar mass loss predictions for each of the SSP templates, allowing a measurement of the current mass and an initial mass at time of formation for each population contained within each spectrum.  Unless otherwise stated, the mass weightings used in this work are the present-day masses of each template, to avoid reliance on the mass loss predictions.  In any case, we find that all results presented here are entirely unaffected by this distinction.

\subsection{Treatment of dust extinction}
\label{sec:FittingDust}

\textsc{Starlight} has the capacity to fit a general dust law with extinction $A_{\textrm{V}}$, and also include an extra extinction Y$A_{\textrm{V}}$ which is applied only to specified templates in the fit.  This could, for example, allow for the possibility that the youngest stellar components are be affected by dust extinction to a greater extent than those populations which would be expected to be free of their birth clouds.  The exact values of Y$A_{\textrm{V}}$ measured by \textsc{Starlight} would in that case be an interesting parameter to model and investigate.  However, in practice, we found that this extra degree of freedom caused \textsc{Starlight}'s fits to be drawn towards negative extinctions when we included a Y$A_{\textrm{V}}$ term for all populations younger than $10^{7.05}$~years.  This is likely due to the combination of the limited wavelength range for which these youngest templates dominate the spectrum due to their extreme colours, and the lack of any significant spectral information beyond their continuum shape.  To the best of our knowledge, the Y$A_{\textrm{V}}$ parameter in \textsc{Starlight} has not successfully been applied to any real spectral fitting to date.

We therefore include a single \citet{Calzetti+00} dust law in the fit, which has the same $A_{\textrm{V}}$ for all templates.  We allow $A_{\textrm{V}}$ to vary in the range of $-1 \leq A_{\textrm{V}} \leq 8$, and we find that over 90\% of the spaxel fits are within the range $0.1 \leq A_{\textrm{V}} \leq 0.8$.

\subsection{Kinematics}
\label{sec:FittingKinematics}

We use the stellar velocity dispersion $\sigma_{\star}$ measured by the MaNGA DAP \citep{DAP} as an initial kinematic guess for the de-redshifted input spectrum's \textsc{Starlight} fits, but allow this to vary as a free parameter in the range of $\sigma_{\star}=20\textrm{ to }900$~km/s.  Unlike other spectral fitting tools such as \texttt{pPXF} \citep{pPXF}, \textsc{Starlight} is not fine-tuned for measuring stellar kinematics, and we do not expect \textsc{Starlight}'s exact measurements of $\sigma_{\star}$ to impact the measured SFHs.  Similarly, despite de-redshifting each spectrum individually using the DAP's stellar velocity $v_{\star}$ measurements, this is a free parameter in the range of $v_{\star}=-600\textrm{ to }600$~km/s to allow \textsc{Starlight} to find its best possible fit, using $v_{\star}=0$ as the initial guess.  In practice, we find that \textsc{Starlight}'s fits are consistent with $v_{\star}=0$, with little deviation in $\sigma_{\star}$ from the DAP measurements.

In the tests outlined in \S\ref{sec:FittingSN} and described in more detail in Appendix~\ref{app:FittingTests}, we find that setting these kinematic parameters to be fixed or variable has no effect on \textsc{Starlight}'s ability to measure stellar populations.  Therefore, to accommodate for any wavelength calibration offsets between the DAP and the SSP templates --- and for any uncertainties in the measurement of the spectral resolution --- we allow the values of $\sigma_{\star}$ and $v_{\star}$ in the fit to vary.

\subsection{Ignoring the youngest stellar populations}
\label{sec:FittingYoung}

We do not expect the star formation rate or chemical evolution to vary significantly within the last $10^{8}$~years in the majority of cases (see e.g. \citealt{SchonrichBinney09}), but initial tests with \textsc{Starlight} revealed that there was a significant correlation between the weights assigned to templates of $\gtrapprox~10^{9.5}$~years and those of $\lessapprox~10^{7.2}$~years, resulting in a sharp peak in the SFH at $\sim~10^{7}$~years.  This effect was found to be present in all locations of all galaxies regardless of signal-to-noise or the strength of dust extinction, and often resulted in an implied SFR of the galaxy to be at least an order of magnitude greater than at any previous time in its history, of up to $\sim~25~M_{\odot}$/yr.  \citet{CidFernandes+GonzalezDelgado10} showed that this phenomenon seems to be related to the known ``UV upturn'' seen in old stellar populations, which is normally attributed to horizontal branch stars in the planetary nebula phase; see \citet{Yi08} for a review.  The cause and presence of this excess of blue light is not accounted for in the old SSP template spectra, so \textsc{Starlight} is forced to attribute it to another population.

We first attempted to mitigate this effect by fitting only from 3700~\AA\ instead of 3541.4~\AA, but found this had no effect on the derived star-formation histories, and we chose not to increase this lower wavelength limit further to avoid impacting the valuable Balmer absorption series.  We then performed another fit with \textsc{Starlight} but where we had combined all of the templates younger than $10^{7.5}$~years for each metallicity into a single template respresenting a flat SFR over that time interval, and used these two templates in the fit instead of the original eight over this time interval.  Comparing the \textsc{Starlight} results of the two approaches shows that enforcing a flat SFR in the youngest templates has no noticeable effect on the SFH in ages $\geq~10^{7.5}$~years at all ($<~0.1$~dex change in the measured SFR at any lookback time $t\geq~10^{7.5}$~years), but the weights assigned to these new templates still exhibited correlation with those assigned to older stellar populations.  Similarly, when we compared \textsc{Starlight} fits using only those SSP templates of ages older than $10^{7.5}$~years, we found that the excess of hot stars was simply assigned to whichever stellar population was youngest.  The rest of the star-formation histories were unaffected, indicating that \textit{older stellar populations are reliably measured regardless of how the youngest populations are treated in the fit}.  We concluded that the youngest stellar population available in the \textsc{Starlight} fit would always have a ``cross-talk'' effect with populations $\gtrapprox~10^{9.5}$~years.  The flux assigned to the youngest populations will always be a combination of the ``true'' flux from stars of that age, as well as a spurious contribution from the hot stars present but not modelled in older populations.

It may be possible to effectively separate these two effects when stellar population models are able to fully model the hot stellar remnants or other factors responsible for the UV upturn.  However, for the purposes of this work, the weights and fluxes of stellar populations younger than 30~Myr are fundamentally unrealiable, so we include these SSP templates in the fit but then \textit{we ignore these populations entirely} and do not use their weights in deriving the \textsc{Starlight}-measured star formation histories.  The SFHs are not likely to have varied over this time period \citep{SchonrichBinney09}, but such young stellar populations are clearly present in many galaxies, so by including these SSPs in the fit but ignoring their weights in subsequent analysis allows the spectrum to be fully modelled.  Limiting the measurements of the derived SFHs to exclude the region younger than 30~Myr does not limit the results from our analyses.

\smallskip

Based on this, we advise users of \textsc{Starlight} and other stellar population fitting software to carefully consider the effects of attempting to measure star-formation histories to young ages without accounting for the limitations of SSP models to include the UV upturn.  Cautious interpretation of all derived SFHs is essential to determine which parts of a SFH are likely to be correctly measured.  However, we do see that the older populations are almost entirely unaffected however the youngest populations are modelled, so are robustly reliable.

\subsection{Effects of low signal-to-noise}
\label{sec:FittingSN}

Many authors spatially bin neighbouring spaxels of integral-field spectroscopic data to create regions with approximately constant signal-to-noise ratio (SNR) before fitting.  However, since we wish to retain the full spatial information of the stellar populations --- and therefore fit each spectrum independently instead of binning --- we must ensure that the \textsc{Starlight} fits in regions with low SNR are reliable. \citet{Ge+18} showed that \textsc{Starlight} may exhibit bias in the fitting of spectra with low SNR, but \citet{FakeNews} contend that these effects are not significant in most physical applications and with the robust \textsc{Starlight} configuration used here.

In Appendix~\ref{app:FittingTests}, we outline a series of tests to measure the effect of low signal-to-noise ratio on \textsc{Starlight}'s recovered fits, and its ability to recover a stellar population of known age or a star-formation history of known shape using the configuration described above.  We find that combining spectra with a given signal-to-noise ratio and comparing the fit of this combined spectrum with the fits of the spectra it contains, \textsc{Starlight} is consistent for the low signal-to-noise regions.  We also find that \textsc{Starlight} is able to recover the age of a known stellar population with a signal-to-noise ratio as low as 5.  Similarly, with the configuration outlined in \S\ref{sec:FittingSLparams}, we find that \textsc{Starlight} can reliably measure the shape of a known SFH in such low signal-to-noise conditions, indicating that we are able to detect the presence of older stellar populations when obscured by brighter younger populations.

These tests imply that, assuming the E-MILES model spectra are accurate representations of the stellar populations they represent, we expect \textsc{Starlight} to be able to recover the true SFHs under all the conditions analysed in the remainder of this paper.  Notwithstanding this robustness, to ensure that the low signal-to-noise regions of the galaxy are not affecting our results in ways we don't anticipate, in all stages of our analysis we ensure that we weight spaxels by their flux or mass, ensuring that the central regions with good fits are up-weighted, and low signal-to-noise regions are down-weighted.

\section{Time-slicing}
\label{sec:AgeSlicing}

From the SSP template weights obtained in the \textsc{Starlight} fits, we are able to reconstruct the star formation history (SFH) and metallicity distributions at every location in each galaxy in the spiral sample.  From the SFHs, it is straightforward to reconstruct an image of the total flux (or mass) emitted by (or contained in) stars of any given age.  To ensure that we are not over-interpreting small-scale noise in the age-distributions of weights assigned to individual templates, we first smooth the SFHs before any analysis is done on these images.  We have smoothed by 0.3~dex in age, but smoothing by any factor between 0.2 and 0.5~dex does not affect results significantly\footnote{We find that the measured uncertainties in the SFHs in the tests described in Appendix~\ref{app:FittingTests} are typically lower than 0.2~dex, but those measurements don't account for any error or bias in the SSPs themselves, so we have conservatively included 0.3~dex smoothing on temporal scales here.}.

\begin{figure}
    \centering
    \animategraphics[width=0.9\linewidth, autoplay, loop, controls, trim=0in 0.5in 0.3in 1in]{15}{animations/galaxy/frames_}{0}{179}
    \caption{\textit{Top}: Animation of a MaNGA spiral (plate-IFU 8329-12701) showing the spatially resolved flux (colour-coded by the metallicity) of stars as a function of age, from 10 to 0.03~Gyr. \textit{Middle}: Weighting function used. The \textsc{Starlight} output's temporal information is smoothed to 0.3~dex. Red points indicate the SSP ages used. \textit{Bottom}: Colour map indicating the flux (in units of $10^{-14}$~erg~s$^{-1}$~cm$^{-2}$~\AA$^{-1}$~spaxel$^{-1}$) and metallicity (in units of log($Z$/Z$_{\odot}$)) of the stellar population. Dashed vertical lines indicate the SSP metallicities used. \textit{Requires Adobe Reader version~$\geq$~9 (and $\leq$~9.4.1 on Linux) or similar. A high-resolution version is available online as supplementary material.}}
    \label{fig:Animation}
\end{figure}

As an illustration, Figure~\ref{fig:Animation} shows an animation of a single galaxy (MaNGA plate-IFU 8329-12701) from the spiral sample, stepping through stellar population ages from 17~Gyr down to 30~Myr, highlighting the wealth of information contained in the spatially-resolved SFHs available using \textsc{Starlight} and MaNGA.  Such animations can be made for any of the galaxies in the sample, but here we show an example of a galaxy observed using the largest-sized (127-fibre) IFU to demonstrate the amount of information potentially available through such time~slicing.

It is worth emphasising that we can only measure the current location of stars in the galaxy, so that we can only treat a ``time~slice'' at any given stellar age as an approximation of the structure of the galaxy at that time, since we cannot undo the effects of dynamical heating or radial mixing and migration.  However, \citet{Martinez-Lombilla+19} showed that the shape of vertical colour gradients seen in edge-on disk galaxies imply that radial migration occurs at a slower rate than the intrinsic growth of the galactic disk.  Simulations of galactic disks also suggest that stellar populations are in general equally likely to migrate inwards or outwards \citep{Avila-Reese+18}, and only by sufficiently small distances that this effect has only minor effects on the radial distribution of populations \citep{Avila-Reese+18, Navarro+18, Barros+20}, so here we assume that the current distribution of a given stellar population is --- to a first approximation --- representative of the distribution of star formation in the galaxy at the corresponding lookback time.

Clearly this assumption does not hold true for spiral structures, since such distributions will become diluted rapidly with the disk's rotation.  However, for the youngest stellar populations, we showed in \citet{Peterken+19TS} that interpreting spatially-resolved star-formation histories in this ``time-slicing'' approach can help to understand spiral arms and bars.  \citet{Mallmann+18} also showed that a similar approach can be used to understand the properties of AGN, and other studies with CALIFA showed that this approach can offer clues to the history of a galaxy's radial profile \citep{CidFernandes+13, CidFernandes+14, Perez+13, GonzalezDelgado+14}.

\section{Mean ages and metallicities}
\label{sec:MeanAgeMets}

A first-order measurement of the SFHs resolved across the face of a galaxy is that of the mean age or --- with a similar calculation --- of spatially-resolved metallicity.  Using the mass weights assigned to each SSP template by \textsc{Starlight} in the fits for each spaxel spectrum, we derive mass-weighted mean age and metallicity (specifically $\langle \log(age/\textrm{yr})\rangle_{\rm{mass}}$ and $\langle \log(Z/Z_{\odot})\rangle_{\rm{mass}}$ respectively) maps.  We then plot the light-weighted median of all spaxels' mean log(age) and log(metallicity) within radial bins of width 0.045~$R_\textrm{e}$ (where $R_\textrm{e}$ is the elliptical Petrosian effective radius measurements from the NSA) against the elliptical galactocentric radius $R$ (in units of $R_\textrm{e}$), and find a best-fit straight line to these data using a least-squares fit.  The fitting is only performed out to 1.2~$R_\textrm{e}$ to avoid the edges of the hexagonal-shaped IFU FOVs and to ensure consistency between galaxies.  From these best-fit lines, we obtain a mean age and metallicity gradient, and a characteristic age and metallicity value of the stellar populations located at 1~$R_\textrm{e}$, a measure which \citet{Sanchez+16} showed to be representative of the galaxy as a whole.

\begin{figure}
    \centering
    \includegraphics[width=\columnwidth]{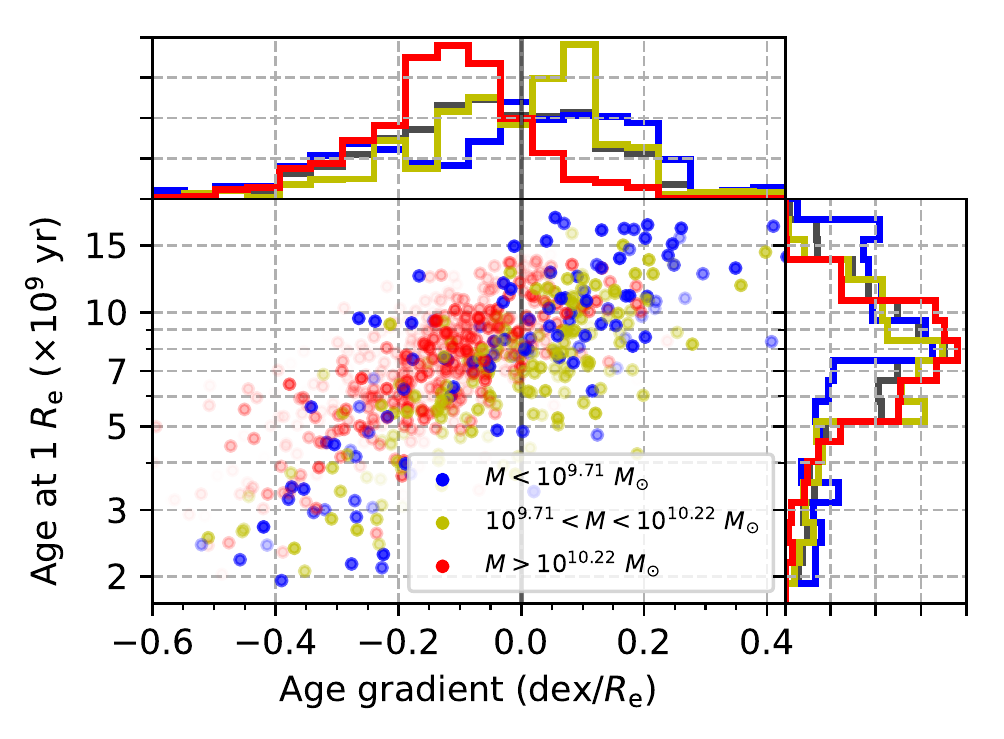}
    \caption{Mass-weighted mean ages at 1~$R_\textrm{e}$ (\textit{vertical axis}, where $R_\textrm{e}$ is the elliptical Petrosian radius measurements from the NSA) and mean age gradients (\textit{horizontal axis}) for each galaxy, coloured by the galaxy's total stellar mass.  Each data point's transparency is defined by the MaNGA Primary+ sample weighting.  The weighted histograms at the \textit{top} and \textit{right} indicate the distributions of ages and their gradients respectively, where the grey line indicates the distribution of the whole sample and the coloured lines indicate those for each of the three mass bins.}
    \label{fig:MeanAges}
\end{figure}

\begin{figure}
    \centering
    \includegraphics[width=\columnwidth]{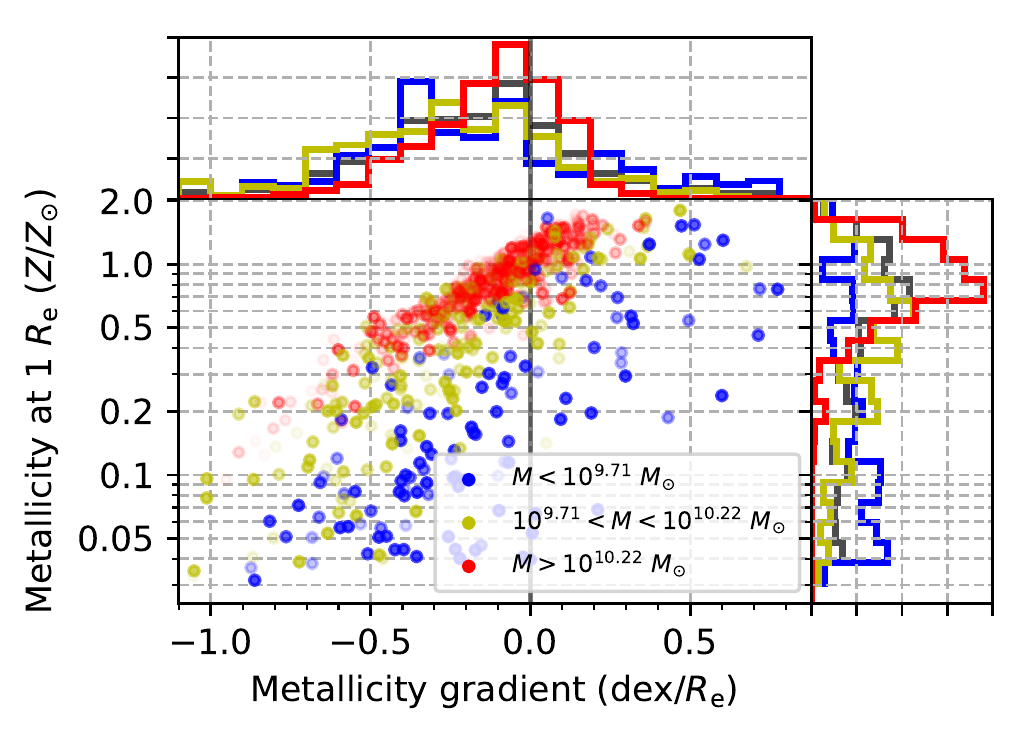}
    \caption{As Figure~\ref{fig:MeanAges}, but for mass-weighted mean metallicities and their gradients.}
    \label{fig:MeanMets}
\end{figure}

The distributions of age gradients and ages at 1~$R_\textrm{e}$ are shown in Figure~\ref{fig:MeanAges}, and equivalent metallicity measurement in Figure~\ref{fig:MeanMets}.  We find that, on average, a majority (approximately 60\%) of the spiral sample exhibit slight negative age gradients, implying younger outskirts.  This agrees with the general picture found by others \citep{SanchezBlazquez+14, GonzalezDelgado+15, Zheng+17, Goddard+17} and is usually taken to be evidence for inside-out formation being dominant in the most massive galaxies.  When the sample is split into three mass bins (of $M<10^{9.71}$~$M_{\odot}$, $10^{9.71}<M<10^{10.22}$~$M_{\odot}$, and $M>10^{10.22}$~$M_{\odot}$\footnote{The mass bin thresholds used here were chosen such that a volume-limited sample of spiral galaxies selected in the method described in \S\ref{sec:SampleSelection} would contain equal numbers of galaxies in each bin, determined using the ``EWEIGHT'' sample weighting for the Primary+ MaNGA sample.}), we find that the approximately 80\% of the highest-mass galaxies exhibit negative age gradients while only 50\% of the lowest-mass galaxies galaxies do.  This difference suggests that inside-out formation is more dominant in high-mass galaxies.

We find that most ($\approx60 - 80$\%) galaxies in all mass bins exhibit slight negative metallicity gradients, and Figure~\ref{fig:MeanMets} highlights a strong mass-metallicity correlation too, as first suggested by \citet{Lequeux+79}.

\section{Mass buildup times}
\label{sec:MassBuildupTime}

\begin{figure}
    \centering
    \includegraphics[width=0.9\columnwidth]{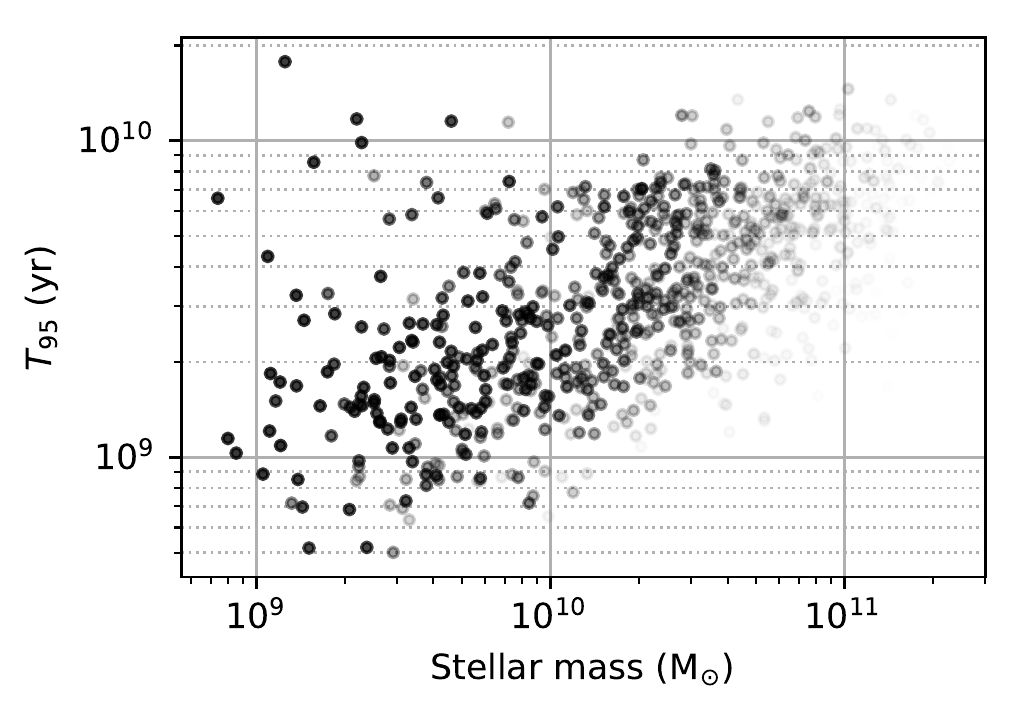}
    \caption{Time since 95\% of the total stellar mass within 1.2~$R_\textrm{e}$ had been assigned in the \textsc{Starlight} fits ($T_{95}$) for galaxies of different present-day stellar mass.  All spiral galaxies with high present-day mass built the bulk of their mass at early times, but most low-mass galaxies were building their mass more recently.  The transparency of each point is defined by the galaxy's MaNGA Primary+ sample described in \S\ref{sec:DataMaNGA}.}
    \label{fig:75MassTime}
\end{figure}

Measuring only a mass-weighted mean age or metallicity does not make use of all of the available information in the age distribution of SSP template weights.  From a full spectrum fitting approach, it is also possible to use the width of the distribution in stellar age, as well as its mean value.  To this end, from a given smoothed SFH, we define the time $T_{95}$ by which 95\% the total stellar mass of that spectrum was built up.  We measure a $T_{95}$ for all light within $R<1.2$~$R_\textrm{e}$ of each galaxy.  We find that $T_{95}$ correlates with the total stellar mass of the galaxy, as shown in Figure~\ref{fig:75MassTime}: all galaxies with present-day stellar masses within 1.2~$R_\textrm{e}$ of $M_{\star} \geq 2\times10^{10} M_{\odot}$ formed the bulk of their mass at least 5~Gyr ago, while most of those with stellar masses $M_{\star} \leq 10^{10} M_{\odot}$ were still building their mass as recently as $\approx 2$~Gyr ago.  This effect is reflected in the known relation between the stellar mass and star formation rates in galaxies, and the results shown here agree well with other fossil record studies \citep{Thomas+10, Pacifici+16}, empirical modelling \citep{Rodriguez-Puebla+17, Behroozi+19}, and theoretical modelling \citep{Henriques+15, Hill+17} including previous analysis of MaNGA galaxies \citep{IbarraMedel+16}.  There is a population of low-mass spiral galaxies with large values of $T_{95}$, but no equivalent population of high-mass galaxies with small build-up times, highlighting that low-mass spiral galaxies have had more varied histories than their high-mass counterparts, as found by \citet{IbarraMedel+16}.

\begin{figure}
    \centering
    \includegraphics[width=0.9\columnwidth]{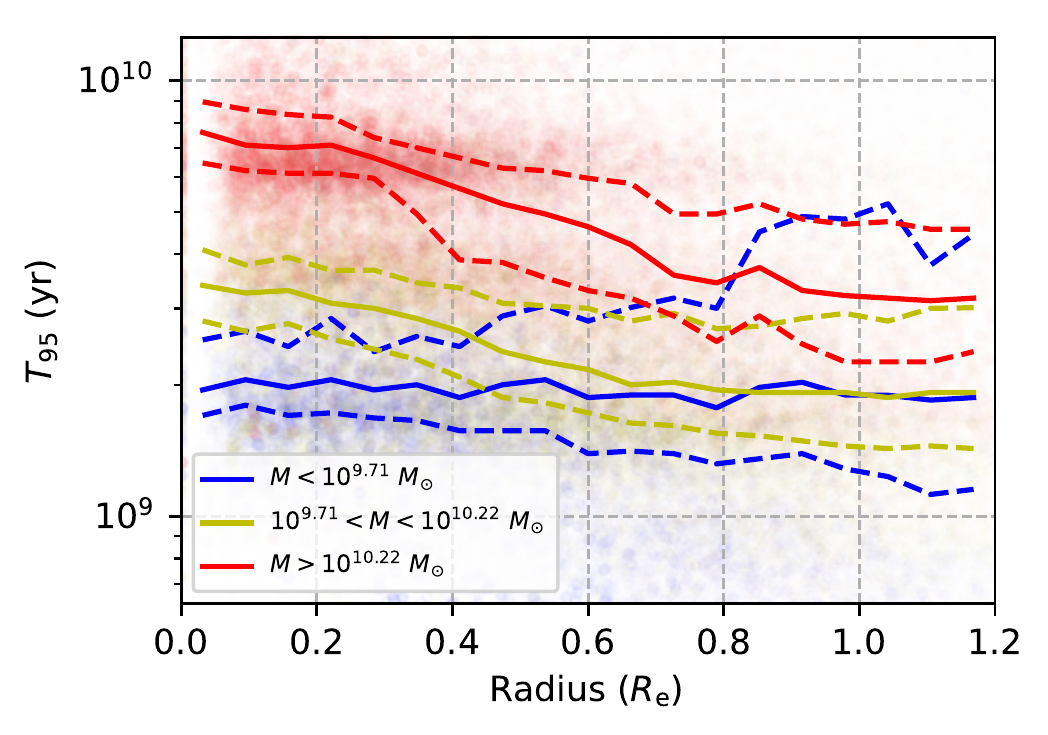}
    \caption{Time since 95\% of the stellar mass built up ($T_{95}$) in each spaxel of each galaxy in the spiral sample.  Each spaxel in each galaxy is shown as a point, with the opacity defined by the product of the total spaxel flux and the galaxy's MaNGA Primary+ sample weight.  Colours denote the galaxy's total present-day mass.  Solid lines represent a weighted running median, and dashed lines are one-third and two-third weighted percentiles.  The outskirts of galaxies of all masses built up at approximately similar times, but the centres of massive galaxies formed significantly earlier than those of low-mass galaxies.  The apparent horizontal feature in the high-mass data points at $\sim7$~Gyr is an artefact: there's a large number of spaxels which have not reached $T_{95}$ by 8.9~Gyr but have by the next oldest SSP at 4.5~Gyr, causing an apparent cluster between these two ages.}
    \label{fig:InsideOutProfiles}
\end{figure}

\begin{figure}
    \centering
    \includegraphics[width=0.9\columnwidth]{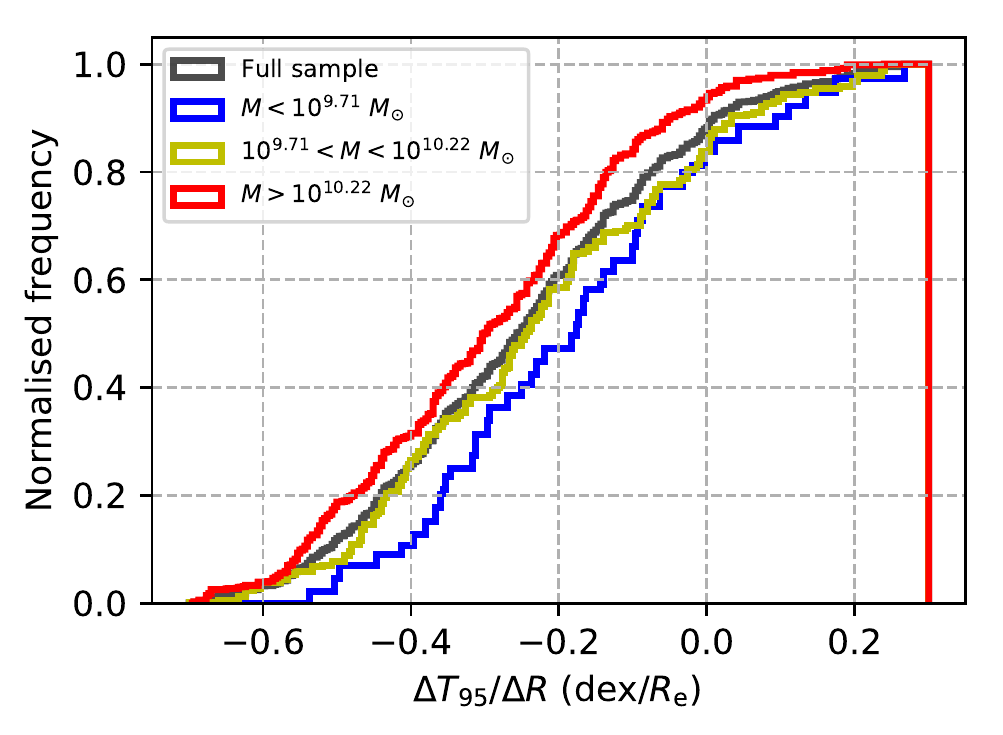}
    \caption{Distribution of gradients of $T_{95}$ vs. galactic radius $R$ for galaxies of different masses.  Most galaxies show evidence for inside-out formation, and the effect is strongest in high-mass galaxies.}
    \label{fig:InsideOutHistogram}
\end{figure}

Using the spatial information available with MaNGA, we are also able to measure how the local value of $T_{95}$ varies with galactic radius $R$ in galaxies of different masses, using the same total stellar mass bins as in Figures~\ref{fig:MeanAges} and \ref{fig:MeanMets}.  In Figure~\ref{fig:InsideOutProfiles}, $T_{95}$ for each spaxel in the sample of spiral galaxies plotted against the galactocentric radius shows that the stellar populations currently at the centres of high-mass galaxies formed on average significantly earlier (by $\approx$~0.7~dex or a factor of 5) than those in low-mass galaxies.  By contrast, the galaxy's outskirts built up at approximately the same time regardless of the mass of the host galaxy.  At $\approx$~1~$R_\textrm{e}$, the discrepancy in $T_{95}$ is much less, at $\approx$~0.3~dex (or a factor of 2).

To quantify this effect, we obtained the radial profiles of $T_{95}$ for each individual galaxy.  We find that these profiles are well-described by a straight line in radius vs.\ $\log(T_{95})$, so we calculate a best-fit straight line, weighting spaxels by their flux.  We find that the majority of galaxies ($>80$\%) in each mass bin show a negative gradient, as we show in Figure~\ref{fig:InsideOutHistogram}, implying younger outskirts than galactic centres.  Assuming that the stellar populations of any given age have not significantly migrated since their birth, this is evidence for inside-out growth occurring in the great majority of spiral galaxies.  We find strongest evidence in the highest-mass galaxies, for which $>$~90\% exhibit negative gradients in $T_{95}$.  These results are consistent with the mean age gradient analysis of \S\ref{sec:MeanAgeMets}, which is not surprising since both approaches are measures of the age distributions contained within the derived SFHs.  However, directly determining a quantity such as $T_{95}$ is returning something much closer to a physical measurement of how the mass of the galaxy has built up over time.

When a 90\%, 75\% or 50\% threshold was used instead of the 95\% threshold results shown here, we found no change to the qualitative conclusions.  The higher 95\% threshold was used to ensure that the buildup time of more galaxies and spaxels was within the range $0.8\gtrapprox T_{95}\gtrapprox 5$~Gyr where spectral fitting methods are most sensitive, and avoids saturation at either extreme of the stellar age range we are able to measure.

\section{Concentration of stellar components}
\label{sec:Concentrations}

Another more physically-motivated way to expand beyond measuring mean age gradients to infer the radial build-up of spiral galaxies is to analyse the spatial extent of individual stellar populations of different ages.  We showed in \citet{Peterken+19TS} that it is possible to measure such distributions directly using time-slicing techniques.  The animation in Figure~\ref{fig:Animation} suggests systematic variation in how concentrated the stellar populations are in one particular spiral galaxy.  Older populations are most centrally-concentrated in the bulge regions of the galaxy while the younger populations make up the more extended disk.  This illustrates the general consensus of the cores of galaxies having younger ages than the surrounding disks.

To quantify the variation in spatial extent of different stellar populations in the full galaxy sample, we choose to measure a concentration of each stellar population in each spiral galaxy.  A concentration can be defined in a number of ways [for example as defined by \citep{Conselice03}] which often require a larger FOV than MaNGA offers in order to measure a background flux.  Here we define the concentration $c$ of a population of stellar age $t$ as
\begin{equation}
    c(t) = \frac{\left<m\right>_{r\leq0.5\textrm{R}_{\textrm{e}}}(t)}{\left<m\right>_{r\leq1.2\textrm{R}_{\textrm{e}}}(t)}
    \label{eq:Concentration}
\end{equation}
where $\left<m\right>_{r\leq k\textrm{R}_{e}}(t)$ is the mean mass contained in all spaxels within $k\times\textrm{R}_{\textrm{e}}$ using the $R_{\textrm{e}}$ elliptical Petrosian radius values of each galaxy from the NASA-Sloan Atlas \citep{NSA}.  This measure ensures that the extent of each population is scaled by the size of the present-day galaxy, and only requires data from within the MaNGA footprint.  This definition of $c(t)$ also means that a completely uniform (i.e.\ radially flat) distribution has a value of $c=1$, with $c<1$ indicating a distribution which rises with radius in the inner region.

\begin{figure}
    \centering
    \includegraphics[width=0.9\columnwidth]{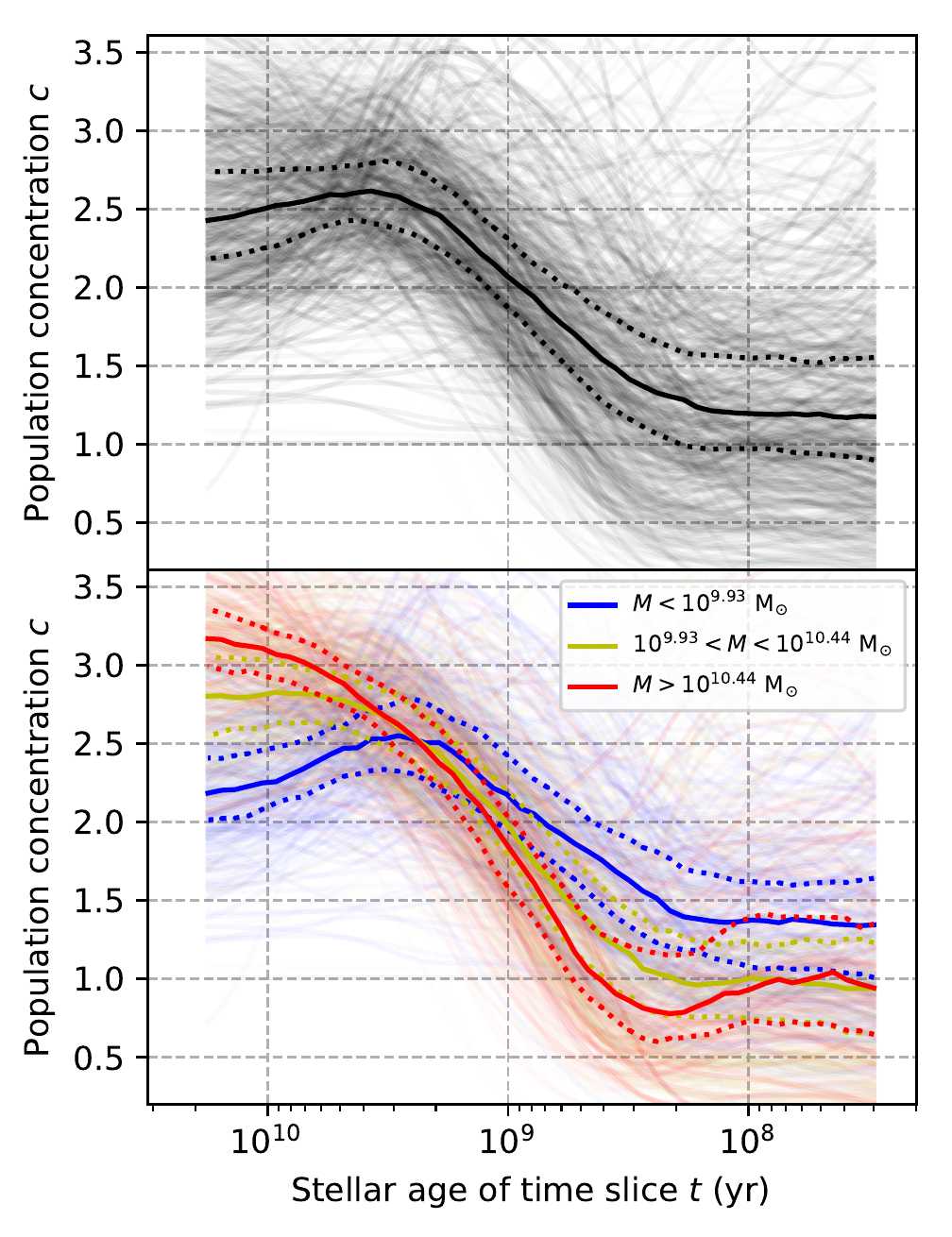}
    \caption{Population concentration $c$ of each galaxy's stellar populations in the spiral sample as a function of stellar age, where $c$ at each time slice is defined by Equation~\ref{eq:Concentration}.  Each galaxy's line is weighted by its MaNGA Primary+ weighting.  The heavy line shows the weighted median of all galaxies, and the dashed lines indicate the weighted one-third and two-third percentiles.  \textit{Top}: All galaxies.  \textit{Bottom}: The same, but with galaxies coloured by their total (present-day) stellar mass.  The youngest stellar populations are more spatially extended than the oldest populations in all galaxies, with the effect strongest in higher-mass galaxies.}
    \label{fig:Concentration}
\end{figure}

The concentration $c$ of stellar populations of different ages in each galaxy in the full sample is shown in Figure~\ref{fig:Concentration}.  There is a clear trend of older populations being most centrally-concentrated (with typical values of $c\approx2.5$ at $t\geq2$~Gyr), and the younger stars in all galaxies exhibiting the most spatially extended distributions (with $c\approx1$ at $t\leq0.1$~Gyr).  This is unsurprising since this is simply a different way of presenting and interpreting the same effects as in \S\ref{sec:MassBuildupTime}, but in a manner that utilises more of the temporal information available to illustrate how radial gradients in mass-to-light ratios (e.g.\ \citealt{GarciaBenito+19}) are created.

We find that there is a strong dependence of $c(t)$ on total (current) galactic stellar mass.  Using the same mass bins as in Figures~\ref{fig:MeanAges} and \ref{fig:MeanMets}, we find that in the highest-mass galaxies, the oldest ($\gtrapprox6$~Gyr) stellar populations are almost three times more concentrated than the youngest populations ($\lessapprox0.1$~Gyr), while in the lowest-mass galaxies this ratio is less than two.

By repeating this analysis using the mean 4020~\AA\ flux mass in the definition of $c(t)$ (i.e.\ replacing $m_{r\leq kR_{\textrm{e}}}(t)$ with $f_{r\leq kR_{\textrm{e}}}(t)$) in Equation~\ref{eq:Concentration}), the results are unchanged.  This is unsurprising since the radial variation in mass-to-light ratio is unlikely to be significant for any single time slice $t$.

This analysis reenforces the conclusion that inside-out growth is the primary formation mode in the majority of spiral galaxies, and that the effect is strongest in higher-mass galaxies.

\section{Mass--size distribution}
\label{sec:MassSize}

Although the mass buildup times in \S\ref{sec:MassBuildupTime} and the variation in concentration in \S\ref{sec:Concentrations} both show evidence for inside-out formation being the dominant growth mechanism in spiral galaxies, these analyses are still not directly comparable measurements to those used in most studies over different redshifts.  Previously, observational evidence for inside-out formation in galaxies has come from analysing how the masses and sizes of galaxies increase simultaneously over time, by measuring these properties of different populations at different redshifts (e.g.\ \citealt{Maltby+10, vanDokkum+13, Patel+13, Papovich+15, Whitney+19}).  This comparison is something that can be directly made using time-slicing methods with integral field spectroscopy for a single galaxy population, to understand how the total mass and size growth has occurred over time.

\begin{figure}
    \centering
    \includegraphics[width=\columnwidth]{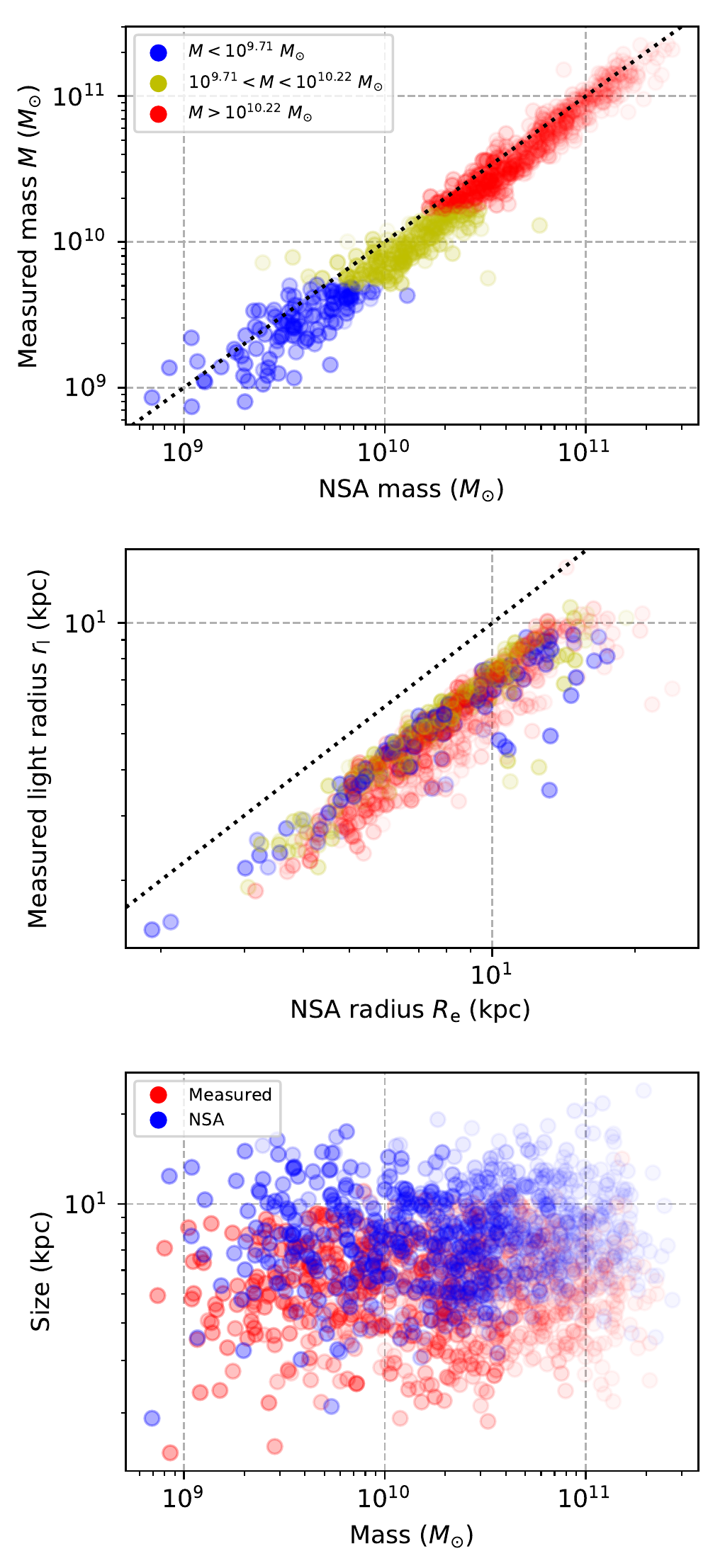}
    \caption{Comparison of the masses (\textit{top}) and radii $r_{\textrm{l}}$ (\textit{middle}) of the spiral sample taken from the NASA-Sloan Atlas \citep{NSA} and measured by \textsc{Starlight} within 1.2~$R_{\textrm{e}}$.  The dashed lines indicate equality.  Both measurements are consistent with the NSA values, with the offset in measured radius $r_{\textrm{l}}$ attributable to the limitations of the MaNGA FOV.  \textit{Bottom}: We find no mass--size relation for the sample of spiral galaxies at the present day when using the \textsc{Starlight}- or NSA-measured parameters (red and blue respectively).  The transparency of the points indicate the relative Primary+ MaNGA sample weighting for each galaxy.}
    \label{fig:MassSizeComparison}
\end{figure}

\subsection{Deriving half-light and mass measurements}
\label{sec:MassSize:measures}

At each stellar age $t$, we define the stellar mass to be the sum of the masses in all populations with ages $\geq t$ within 1.2~$R_{\textrm{e}}$, using the temporally-smoothed distribution of weights from \textsc{Starlight}.  We can also define a measurement $r_{\textrm{l}}(t)$ of the light size of a time slice $t$ as being the radius of half the light contained within 1.2~$R_{\textrm{e}}$ (using $R_{\textrm{e}}$ elliptical Petrosian radius measurements from the NSA) of all of the light emitted by stars older than $t$.  This definition is used since the MaNGA observations are limited in their fields of view.  This limitation prohibits us from reliably measuring a sky background, forbidding a direct half-light radius measurement in a normal approach.

To ensure that the radius and stellar mass measurements defined here using the \textsc{Starlight} fits are reliable, we compare these measurements for the present-day galaxy (i.e.\ $t=0$) with the known size and mass measurements of the galaxies in the NSA (see Figure~\ref{fig:MassSizeComparison}).  We find that $r_{\textrm{l}}$ is a good proxy for the NSA elliptical Petrosian half-light measurements, with an offset of $\approx 0.2$~dex which is a consequence of both the limited MaNGA FOV and the difference in wavelengths used.  (The NSA radii are measured in the \textit{r} band imagery, but the measurements for the \textsc{Starlight} outputs are done on a model 4020~\AA\ image, which would be located in the \textit{g} band.)  We also find that the total stellar masses determined by \textsc{Starlight} are consistent with the photometry-derived masses in the NSA.  Both mass measurements assume the same IMF, so a small observed offset is likely due to MaNGA's limited FOV.

\citet{Maltby+10}, \citet{Mosleh+13}, \citet{Bernardi+14}, \citet{vanderWel+14} and others have shown that, unlike the early-type galaxies, the mass--size relation for spiral galaxies is weak.  However, using the \textsc{Starlight}-derived measurements of the galaxies' masses and sizes, we find no strong mass--size trend in the present day sample of spiral galaxies at all; a Spearman rank test results in a correlation p-value of only $p=0.84$ for the measured data, and similar for the NSA values.  This lack of a relation may indicate that the Galaxy~Zoo classifications for low-mass galaxies under the conservative selection criteria used here may be slightly biased so that the smaller low-mass galaxies are less likely to be classified as spirals.

\begin{figure*}
    \centering
    \animategraphics[width=\textwidth, autoplay, loop, controls, trim=0in 0.3in 0.3in 1in]{15}{animations/MS/frames_}{0}{169}
    \caption{Evolution of the mass--size plane of the spiral sample over time.  The lookback time and corresponding redshift are indicated at the top of the figure.  The left column shows the mass--size plane for the light radius ($r_{\textrm{l}}$, \textit{top}) and mass radius ($r_{\textrm{m}}$, \textit{bottom}) measurements.  The right column indicates the overall change in each galaxy's mass and size (in dex) from the first frame of the animation.  The redshift of each galaxy is accounted for, such that in any given frame the star-formation histories of each galaxy is sampled at the difference between the frame age and the lookback time implied by the galaxy's redshift.}
    \label{fig:MassSizeAnimation}
\end{figure*}

\begin{figure*}
    \centering
    \includegraphics[width=\textwidth]{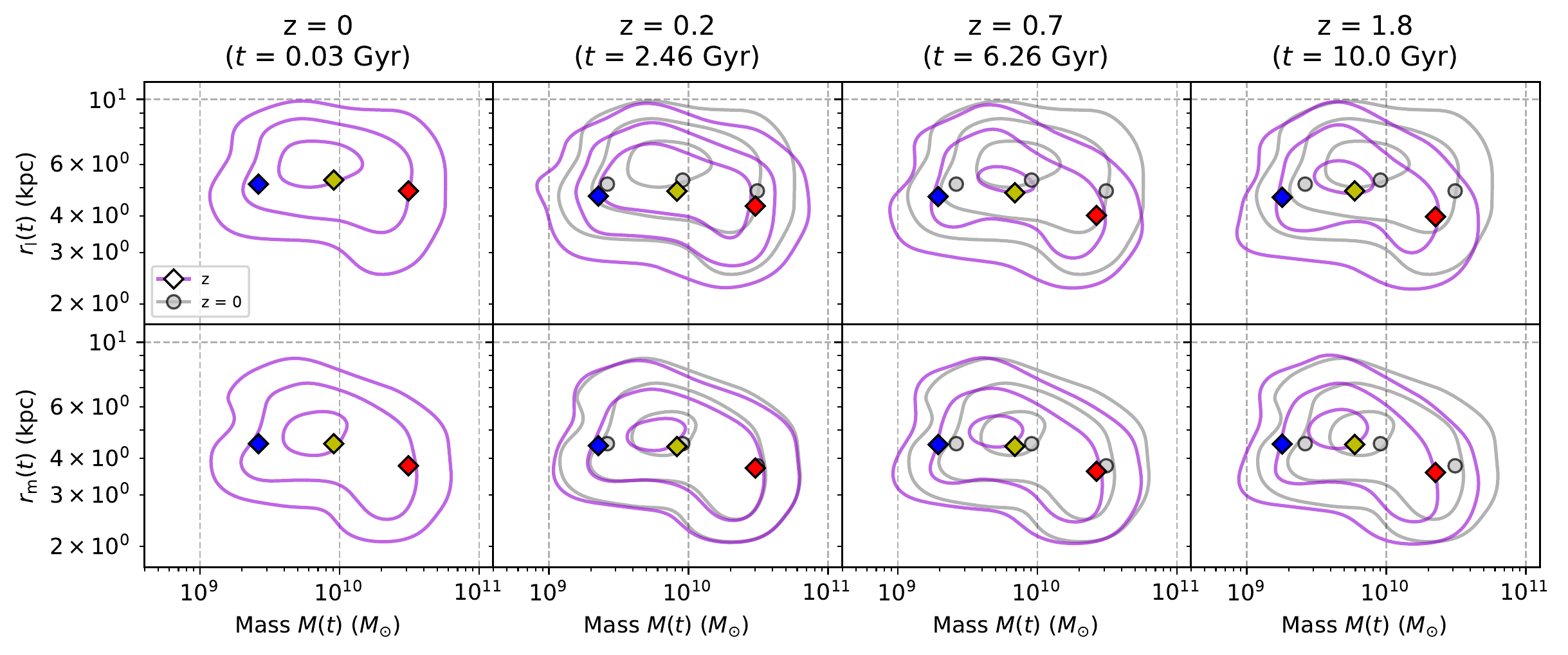}
    \caption{The distributions of galaxies at individual snapshots in Figure~\ref{fig:MassSizeAnimation} showing the mass--$r_{\textrm{l}}$ (\textit{upper panels}) and mass--$r_{\textrm{m}}$ (\textit{lower panels}) planes at selected redshifts z.  The corresponding lookback times $t$ are also indicated.  As fiducials, the grey contours and circle markers indicate the distribution of galaxies and the mean positions of each mass bin at z~=~0, while the magenta contours and coloured diamond points indicate the distribution and mean positions at each redshift's time slice.}
    \label{fig:MassSizeSnapshots}
\end{figure*}

\subsection{Evolution of the mass--half-light-radius plane}
\label{sec:MassSize:rl}

Having reassured ourselves that our mass and $r_{\textrm{l}}$ radius measurements are appropriate proxies for the photometric measurements in the present-day galaxies, we can now explore how the mass--size plane changes over time.  The upper panels of the animation in Figure~\ref{fig:MassSizeAnimation} shows the evolution of the mass-$r_{\textrm{l}}$ plane over the last $\approx10$~Gyr.  Figure~\ref{fig:MassSizeSnapshots} also shows the distribution of galaxies in the mass--$r_{\textrm{l}}$ plane at four different redshifts.  The measurements shown are using \textsc{Starlight}'s current mass measurements of each SSP template (see \S\ref{sec:FittingSLparams} for the distinction between current and initial mass weights).  In reality, the mass loss of each population will have been gradual over the galaxy's evolution rather than instantaneous as this approach implies.  However, as stated in \S\ref{sec:FittingSLparams}, by instead adopting the initial mass --- and therefore assume that no mass loss occurs at all --- we find no significant change to these results.  The ``reality'' would of course be between these two extremes.  However, since the two cases reach near-identical results, we present here only the results for the current mass template weightings to avoid uncertainties in modelling time-dependent mass loss estimates separately for each SSP at each time-step.

Assuming an absence of significant systematic radial migration effects, we find that the growth in $r_{\textrm{l}}$ of these galaxies has only occurred over the last $\approx 3$~Gyr, while the bulk of the growth in mass occurred before this.  We also find that galaxies generally have not changed their relative mass group, instead growing in mass and size at the same rates as those of similar masses and sizes.  This cohort behaviour implies that, although every galaxy has had a unique formation history, tracing the average evolution of a galaxy population (e.g.\ by measuring galaxy properties at different redshifts) is representative of how most galaxies have evolved over the same time period.

\subsubsection{Mass dependence}
\label{sec:MassSize:rl:mass}

By splitting the galaxy into the subsamples of different mass bins as before, we find that over the last 10~Gyr, the low-mass galaxies have grown significantly more in mass ($\approx 0.17$~dex) but less in light radius $r_{\textrm{l}}$ ($\approx 0.05$~dex) than the high-mass galaxies ($\approx 0.14$~dex growth in mass, $\approx 0.1$~dex in $r_{\textrm{l}}$).  The small $\approx 0.03$~dex difference in mass growth rates between the samples combined with the mass dependence of $T_{95}$ seen in \S\ref{sec:MassBuildupTime} indicates that the low-mass galaxies have only built up slightly more mass relative to the high-mass galaxies, but that this growth occurred later.  We also see this downsizing effect in the ``turnup'' time --- at which galaxies stop growing significantly in mass and start growing in light radius $r_{\textrm{l}}$ --- which occurred earlier in high-mass than low-mass galaxies ($\approx$~3.5~Gyr ago compared to $\approx$~1~Gyr ago).

\subsection{Evolution of the mass--half-mass-radius plane}
\label{sec:MassSize:rm}

While the light-weighted radius measurements are directly comparable to the size evolution of galaxy populations observed at different redshifts, the mass distribution of a galaxy is more fundamental to its build-up.  In Figures~\ref{fig:MassSizeAnimation} and \ref{fig:MassSizeSnapshots}, we therefore also show the evolution of the mass--size plane but using a half-\textit{mass} size $r_{\textrm{m}}$ (equivalently defined as the radius containing half of the stellar mass within 1.2~$R_{\textrm{e}}$ due to the limitations of the MaNGA FOV) using the SSP template mass-to-light ratios.  We find that despite increases in the observed light size $r_{\textrm{l}}$ of the galaxy population, the corresponding increase in the mass size $r_{\textrm{m}}$ of the same galaxies is minimal; we find an increase of $\lessapprox$~0.05~dex in size for almost all galaxies, even in those with low present-day stellar masses.  This weak evolution is in agreement with the results presented by \citet{Suess+19a,Suess+19b} using entirely independent approach to show that the half mass radius does not evolve significantly compared to the evolution of the half light radius.

The physical size growth of spiral galaxies over the last 10~Gyr has therefore been extremely small, at typically only 10\% growth.  Such an increase in mass radius --- however slight --- requires a radial increase in the regions of ongoing star formation.  Since younger stellar populations dominate the light of a spiral galaxy at any time slice or lookback time, the increase in measured radius in observations of the same galaxies becomes significant.  A small amount of star formation in the outskirts of the galaxies will contribute a large amount to the light while contributing comparatively little to the bulk of the galaxy, causing a strong mass-to-light gradient.  Direct measurements of the growth of galaxies from observations therefore produce an overestimate of the underlying mass growth rate.  This effect has also been recently quantified for cosmological galaxy catalogs from CANDELS  \citep{Suess+19a, Suess+19b}, who showed that the half-light radius growth of galaxies, both star-forming and quiescent, previously reported in many works is significantly weaker for the half-mass radius.

Interestingly, it has been reported by \citet{Frankel+19} that the structure of stellar populations seen in the Milky Way provide evidence for a slower growth in half-mass radius than in the half-light radius, and the evidence presented here --- as well as from high-redshift surveys (see above) --- suggests that this feature is common in the growth of spiral galaxies.  This slow size growth of spiral galaxies seems to be in tension with predictions from semi-analytical models and hydrodynamics simulations of galaxy evolution in the context of the $\Lambda$CDM cosmology (see for a discussion \citealt{Avila-Reese+18} and more references therein).

%Instead, we see that spiral galaxies have grown their mass over time without significantly altering their physical size.  The observed apparent increase in light radius therefore only represents a small amount of recent (and therefore bright) star-formation in the galactic disks which does not alter the distributions of total mass of the galaxy, even in galaxies with low present-day stellar masses.

\section{Potential limitations}
\label{sec:Discussion}

\subsection{Limitations of the data}
\label{sec:DataDiscussion}

Due to the limited FOV of the MaNGA observations, we are unable to measure a true half-light (or half-mass) radius for any given ``time slice'', since we do not have any background in the images.  We are able to confirm in Figure~\ref{fig:MassSizeComparison} that the radius of half of the light (or mass) contained within 1.2~$R_{\textrm{e}}$ is a good proxy for the present-day galaxy, but we have no way of confirming this at other stellar population ages.  However, since we find that the oldest populations are most concentrated, the measured sizes in the earlier age-steps in the mass--size evolution are likely to be closer to the true sizes.  The observed increase in size is therefore a conservative estimate of the real change.  Since little of a galaxy's mass is located outside 1~$R_{\textrm{e}}$ (e.g.\ \citealt{Perez+13}), we expect that the mass radius $r_{\textrm{m}}$ measurements are likely to be close to true half-mass radii.

%To accurately measure how the distribution of different stellar populations changes with stellar age in any single galaxy, integral-field spectroscopic observations would be required to a larger galactic radius than is available with MaNGA.  However, the strength of the MaNGA observations is in the sample size.  We have been able to perform a fossil record analysis of an entire population of galaxies, allowing us to extract trends over all spiral galaxies.

\subsection{Stellar population models and spectral fitting}
\label{sec:SSPFittingDiscussion}

Although we show in Appendix~\ref{app:FittingTests} that \textsc{Starlight} can measure stellar populations if the models used to do so are correct, this work assumes that the model spectrum templates of the E-MILES \citep{E-MILES} and \citet{Asa'd+17} libraries are representative of the true observed stellar populations.  There are a number of unresolved problems in the field of stellar population modelling; see \citet{Conroy13} for a comprehensive review.  For example, in \S\ref{sec:FittingYoung}, we described a correlation between weights assigned to populations $\gtrapprox~10^{9.5}$~years and those of $\lessapprox~10^{7.2}$~years due to a deficiency in the SSP templates.  This is likely to be related to the UV upturn problem, due to the presence of hot stars in old stellar populations \citep{Yi08} which is not accounted for in the SSP models.  There is also uncertainty surrounding the shape of the IMF and ongoing debate on whether it varies between and within galaxies \citep{LaBarbera+13, Alton+17, Vaughan+18, Parikh+18}.  In principle, any variation of the IMF over cosmic time is likely to affect our analysis too.

We also assume here that stellar metallicity is a one-dimensional parameter.  In reality, the individual elemental abundances can vary from star to star.  Further time-slicing work can be done to measure the simultaneous change in star-formation histories and metallacity evolution, including variation in $\alpha$-enhanced metals, but this is beyond the scope of this project.  Although we are confident that the fitting methods used here can recover the distributions of stellar population ages and metallicities, the degeneracy between metallicity and [$\alpha$/Fe] is harder to assess.  However, in \citet{Peterken+19TS}, we found that removing the extra metallicity dimension appears to have little effect on the derived star-formation history.

This work has also assumed a single \citet{Calzetti+00} exinction model which affects every stellar population contained within a single spectrum equally.  As we state in \S\ref{sec:Fitting}, we expect that younger stellar populations are instead likely to be affected by a greater amount of extinction, but we are unable to resolve this difference in non-parametric fitting using \textsc{Starlight}.  How this deficiency affects the measured star-formation histories is not known.

Notwithstanding these shortcomings and assumptions used in the fitting process, the resulting star-formation histories tell a consistent story of inside-out formation in spiral galaxies with no noticeable artifacts, and the coherent structures visible in the time-slicing of galaxy 8329-12701 shown in Figure~\ref{fig:Animation} gives confidence in the fitting method for the purposes described here.  The clear inside-out formation reported here might even be underestimated: \citet{IbarraMedel+19} have recently shown that any intrinsic signature of inside-out growth is diminished by the instrumental/observational setting and the stellar population modelling, mainly the age resolution of the SSP templates.

%Additionally, \citet{IbarraMedel+19} show that any instrinsic signatures of inside-out growth is likely to be diluted when measured using stellar population modelling, due to the difficulty in identifying faint populations underlying a much brighter population of a different age.  This suggests that the evidence we find here is not due to any systematic uncertainties from our assumptions and methods and is instead measuring a real imprint of the stellar populations left in the galaxies as they have grown in mass and size.

\subsection{Effects of radial mixing and mergers}
\label{sec:RadMigDiscussion}

Time-slicing methods can only reveal the current locations of different stellar populations in a galaxy.  In this work, we have interpreted these present-day distributions to be indicative of the radial distributions of star formation at the age of the stellar population, and make no attempt to correct for the effects of radial migration or mergers.  Fortunately, simulations suggest that the radial distributions of stellar populations in a galactic disk are not significantly altered by radial migration \citep{Avila-Reese+18, Navarro+18, Barros+20}, indicating that the assumptions made here are at least approximately valid.

%However, if radial migration is significant in the stellar populations analysed here, is is likely to occur in no preferential direction (i.e. inwards or outwards). 

High-resolution simulations of Milky Way-like galaxies show that radial migration has no preferential direction, with most stars being scattered similarly inwards and outwards, by typically no more than 1--2~kpc \citep{Avila-Reese+18}.  Instead, stars are equally likely to move in either direction over their lives \citep{SellwoodBinney02, Avila-Reese+18}, with observations implying that any resulting observed growth as a result of migration occurs slower than the intrinsic growth of the disk \citep{Martinez-Lombilla+19}.  Any radial migration of an initially centrally-concentrated distribution of stars is likely to become slightly less concentrated over time, an effect which is observed in the stellar metallicity distributions of the solar neighbourhood in the Milky Way (e.g.\ \citealt{Feltzing+19, Frankel+18}).  A galaxy with a radial distribution of star formation that is not varying over time would be observed using time-slicing methods to have been slightly decreasing in measured radius over the same time frame, since the oldest populations will have more time to disperse and would therefore appear at larger radii.  The measured variations of spatial distribution of stellar populations of different ages in \S\ref{sec:Concentrations} are also therefore likely to be a close lower limit on the true variation of the sizes of spiral galaxies over the same time period.  Similarly, the recovered change in light size $r_{\textrm{l}}$ in \S\ref{sec:MassSize} is therefore a slightly conservative but representative estimate of how the galaxy evolved over the same time period.

\section{Conclusions}

We have derived spatially-resolved star formation histories for a sample of 795 low-redshift spiral galaxies using \textsc{Starlight} applied to integral-field spectroscopic observations from SDSS-IV MaNGA.  From this fossil record analysis, we have built maps indicating the regions in which stellar populations of different ages are located in any given galaxy.  We analysed the radial profiles of these ``time slices'' to extract the historical growth of the population of spiral galaxies.  The main findings are:

\begin{itemize}
    \item Using E-MILES single stellar population template spectra, the star formation histories measured by \textsc{Starlight} are unreliable for the youngest populations used in the fit (in this case those younger than $3\times10^{7}$~years).  We found evidence that this is related to the UV upturn \citep{Yi08} and a solution to this problem requires population models to include the presence of hot old stars (whatever their nature) in the oldest population templates.  However, despite this degeneracy between the oldest and youngest template weights, the derived star-formation histories of the stellar populations older than $3\times10^{7}$~years are trustworthy.
    % \item Assuming that the E-MILES templates accurately represent the populations they are modelling, \textsc{Starlight} is able to determine the star-formation histories in MaNGA spectra with as low signal-to-noise ratio as 5.
    \item We have quantified evidence for inside-out galaxy growth in three different ways, which all indicate that such a growth mode is dominant in the majority of spiral galaxies, and is most significant in high-mass galaxies:
    \begin{itemize}
        \item The mass-weighted mean age gradient of spiral galaxies tends to be slightly negative; the outskirts are younger than the centres in $\approx 60$\% of all spiral galaxies.  This fraction rises to 80\% for galaxies with stellar mass $M>10^{10.22}\textrm{M}_{\odot}$.
        \item By measuring a time $T_{95}$ by which 95\% of the stellar mass had built up in each location of the galaxy, we find that $T_{95}$ decreases with radius in the majority galaxies.  Gradients in $T_{95}$ are steepest in the highest-mass galaxies.
        \item The concentration $c$ of each ``time slice'' was found for each galaxy.  The youngest stellar populations (younger than $\approx 10^{8.5}$~years) are more radially extended than the oldest ($\approx 10^{10}$~years old) populations in all cases, and this effect is most significant in high-mass galaxies.\end{itemize}
    \item By considering the simultaneous increase in stellar mass and the increase in light radius with the addition of ever-younger stellar populations, we found that the mass--size distribution of spiral galaxies evolves with very little change in rank; galaxies grow in mass and size at similar rates to other galaxies with similar masses and sizes.  This suggests that a ``like for like'' approach when comparing the sizes and masses of distributions galaxies at different redshifts is representative of how the individual galaxies themselves have evolved.
    \item We found that over the last 10~Gyr, galaxies with high present-day stellar masses have grown their half-light size by approximately twice the amount that low-mass galaxies have, although low-mass galaxies have grown slightly more in mass.
    \item However, when the half-mass radius of the galaxies was used instead, we found that spiral galaxies have barely altered their radial mass distributions over the same time period.  Although galaxies appear to grow in (light) size over cosmic time, we show that this is an overestimate of their actual physical growth.  This apparent discrepancy is due to a small amount of star formation occurring in the outskirts being able to dominate a galaxy's light while contributing very little to the physical bulk of the galaxy.
\end{itemize}

\section*{Acknowledgements}

Funding for the Sloan Digital Sky Survey IV has been provided by the Alfred P. Sloan Foundation, the U.S. Department of Energy Office of Science, and the Participating Institutions.  SDSS-IV acknowledges support and resources from the Center for High-Performance Computing at the University of Utah. The SDSS web site is \url{www.sdss.org}.

SDSS-IV is managed by the Astrophysical Research Consortium for the Participating Institutions  of  the  SDSS  Collaboration including the Brazilian Participation Group, the Carnegie Institution for Science, Carnegie Mellon University, the Chilean Participation Group, the French Participation Group, Harvard-Smithsonian Center for Astrophysics, Instituto de Astrof{\'i}sica de Canarias, The Johns Hopkins University, Kavli Institute for the Physics and Mathematics of the Universe (IPMU) / University of Tokyo, Lawrence Berkeley National Laboratory, Leibniz Institut f{\"u}r Astrophysik Potsdam (AIP), Max-Planck-Institut f{\"u}r Astronomie (MPIA Heidelberg), Max-Planck-Institut f{\"u}r Astrophysik (MPA Garching), Max-Planck-Institut f{\"u}r Extraterrestrische Physik (MPE), National Astronomical Observatories of China, New Mexico State University, New York University, University of Notre Dame, Observat{\'o}rio Nacional / MCTI, The Ohio State University,  Pennsylvania State University, Shanghai Astronomical Observatory, United Kingdom Participation Group,  Universidad Nacional Aut{\'o}noma de M{\'e}xico, University of Arizona, University of Colorado Boulder, University of Oxford, University of Portsmouth, University of Utah, University of Virginia, University of Washington, University of Wisconsin, Vanderbilt University, and Yale University.

This publication uses data generated via the Zooniverse.org platform, development of which is funded by generous support, including a Global Impact Award from Google, and by a grant from the Alfred P. Sloan Foundation.

We are grateful for access to the University of Nottingham's \texttt{Augusta} high performance computing facility.

J.H.K. acknowledges financial support from the European Union's Horizon
2020 research and innovation programme under Marie Sk\l odowska-Curie
grant agreement No 721463 to the SUNDIAL ITN network, from the State
Research Agency (AEI) of the Spanish Ministry of Science, Innovation and
Universities (MCIU) and the European Regional Development Fund (FEDER)
under the grant with reference AYA2016-76219-P, from IAC project
P/300724, financed by the Ministry of Science, Innovation and
Universities, through the State Budget and by the Canary Islands
Department of Economy, Knowledge and Employment, through the Regional
Budget of the Autonomous Community, and from the Fundaci\'on BBVA under
its 2017 programme of assistance to scientific research groups, for the
project "Using machine-learning techniques to drag galaxies from the
noise in deep imaging".

%%%%%%%%%%%%%%%%%%%%%%%%%%%%%%%%%%%%%%%%%%%%%%%%%%

%%%%%%%%%%%%%%%%%%%% REFERENCES %%%%%%%%%%%%%%%%%%

% The best way to enter references is to use BibTeX:

\bibliographystyle{mnras}
\bibliography{refs} % if your bibtex file is called example.bib

% Alternatively you could enter them by hand, like this:
% This method is tedious and prone to error if you have lots of references

%%%%%%%%%%%%%%%%%%%%%%%%%%%%%%%%%%%%%%%%%%%%%%%%%%

%%%%%%%%%%%%%%%%% APPENDICES %%%%%%%%%%%%%%%%%%%%%

\appendix

\section{Testing full-spectrum fitting}
\label{app:FittingTests}

It is common to spatially bin neighbouring spaxels of integral-field spectroscopic data to create regions with a minimum signal-to-noise ratio (SNR) before fitting.  However, since we wish to retain and measure the full spatial information of the stellar populations --- and therefore fit each spectrum independently instead of binning --- we require the \textsc{Starlight} fits in regions with low SNR to be reliable. \citet{Ge+18} showed that \textsc{Starlight} may exhibit bias in the fitting of spectra with low SNR, but \citet{FakeNews} contends that these effects are not significant in most physical applications and with the robust \textsc{Starlight} configuration used here.  To assess this conclusion, we have performed a series of tests laid out here.

\subsection{Average fits of regions with low signal-to-noise}
\label{app:FittingTests-SN}

To test how the signal-to-noise ratio (SNR) of a spectrum in a MaNGA datacube may affect the fitting results from \textsc{Starlight}, we combined spectra of a single galaxy (plate-IFU 8329-12701) within different SNR bins to form a single integrated spectrum for each bin.  In combining spectra from the MaNGA datacube, emission lines were removed and the spaxel spectra de-redshifted and interpolated onto a common wavelength base before summing.  Each single spaxel's SNR was then defined as the median value over the fitting wavelength range of the ratio between the spaxel's flux spectrum and the reciprocal of the square root of the inverse variance spectrum (as measured by the MaNGA DRP).

We chose to combine spaxel spectra in SNR bins of width 2, centred on every even value.  A single spectrum was created by combining all spectra from spaxels with SNR between 3 and 5, another from spaxels with SNR between 5 and 7, etc., up to a spectrum comprising the sum of all spaxels with a SNR between 29 and 31.  Each of the combined spectra's signal-to-noise ratio is greater than 60 and most are greater than $\sim$~200.  These combined spectra were then fit using \textsc{Starlight} with an identical configuration to that of the science fitting to see how their measured star-formation histories varied from the average of their constituent parts.  In the absence of any systematic bias in \textsc{Starlight}, it would be expected that the average SFH measured in all individually-fitted spaxels in a given SNR bin should be the same as the SFH measured from the average spectrum of those spaxels, regardless of the actual variation in SFH shape between spaxels in a given bin.

\begin{figure}
    \centering
    \includegraphics[width=\columnwidth]{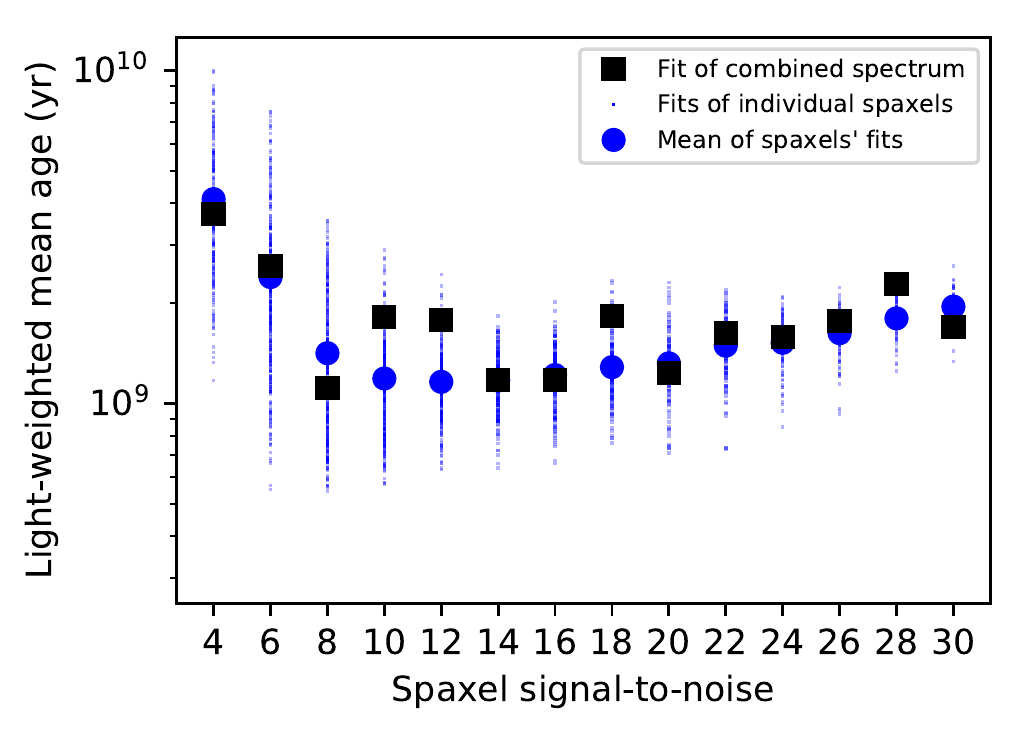}
    \caption{The mean age of spaxels in different signal-to-noise ratio bins (\textit{blue}) in a spiral galaxy compared to the mean age of the spectrum of all spaxels combined (\textit{black squares}).  There is no significant bias in the average age measured by \textsc{Starlight} compared with varying SNR.}
    \label{appfig:AgeRecovery}
\end{figure}

\begin{figure}
    \centering
    \includegraphics[width=\columnwidth]{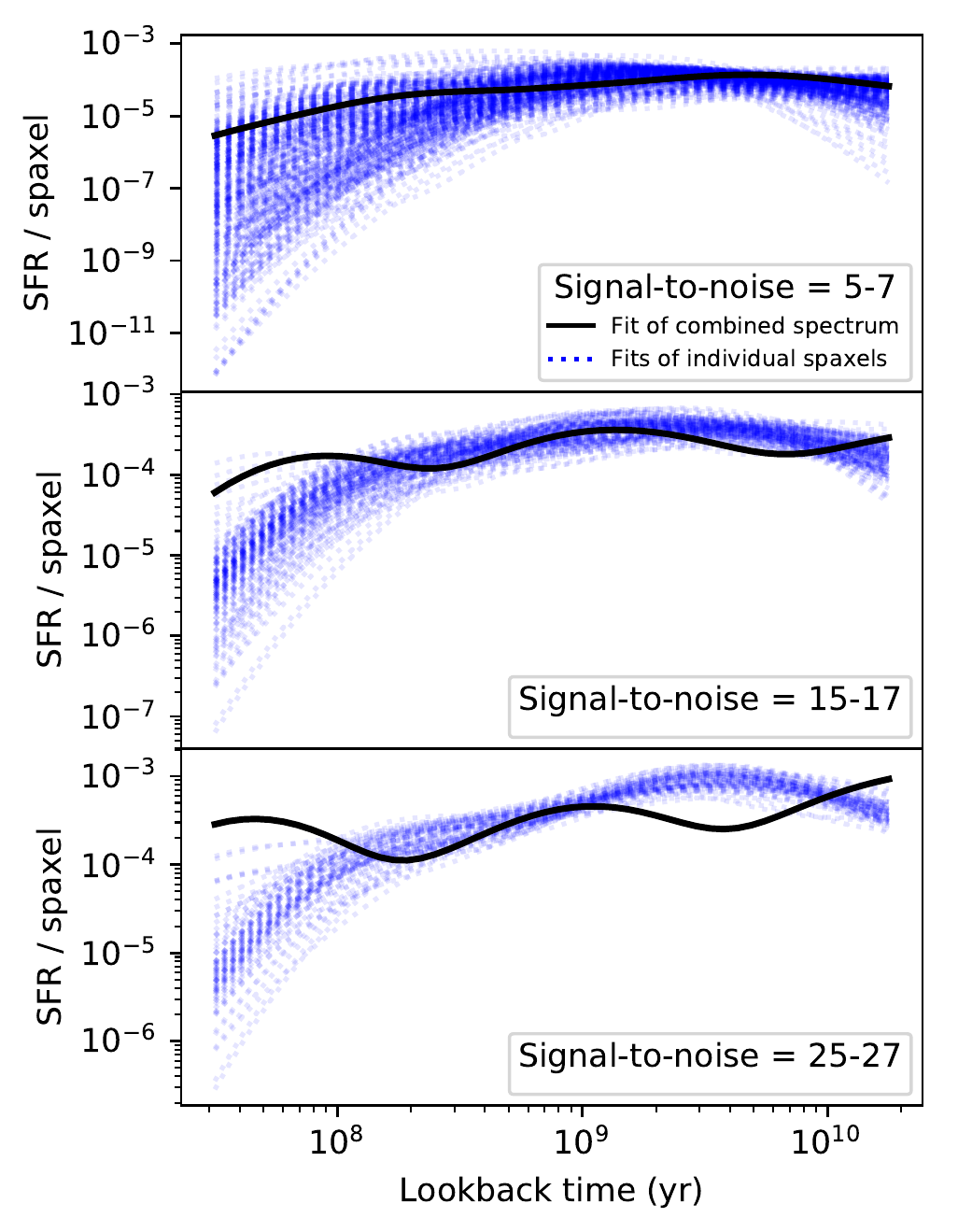}
    \caption{The star-formation histories of spaxels (\textit{blue}) in three different signal-to-noise ratio bins compared to the SFH of the spectrum of all spaxels combined (\textit{black}).  \textsc{Starlight} shows worse performance at higher signal-to-noise ratios.}
    \label{appfig:SFHRecovery}
\end{figure}

We find that the light-weighted mean age of the summed spectrum is always within $\sim0.2$~dex of the mean age of all individual spaxels (separately fitted) in a given SNR bin, indicating that signal-to-noise effects do not significantly bias the average results (see Figure~\ref{appfig:AgeRecovery}).  However, we find that the full star-formation histories of the summed spectra are most discrepant for the bins of larger SNR, as shown in Figure~\ref{appfig:SFHRecovery}.  This is likely due to an effect of small systematics (e.g.\ sky subtraction or flux calibration) dominating over random noise when summing spectra of already-high signal-to-noise ratios.  In summing high-SNR spectra, the modest reduction in combined SNR is outweighed by the increase in systematic errors when considering the fine detail required to measure a SFH.  The fact that these effects have less significance in measuring the average properties highlights the level of extra complexity involved in measuring SFHs over mean ages.

\subsection{Recovery of a single stellar population of known age}
\label{app:FittingTests-SSPRecovery}

To further examine whether the above effects are due to systematics in the spectra rather than in \textsc{Starlight}, we tested how well \textsc{Starlight} is able to return the age and metallicity of a single stellar population with known parameters.  We can create spectra representing single stellar populations of any age and metallicity by interpolating over the grid of E-MILES SSP template spectra.  We produced spectra representing 200 ages and three metallicities using a bilinear interpolation in 2D log space of the 66 SSPs used in the fitting.  We then degraded these spectra to signal-to-noise ratios of 5, 10, 15, and 20 by adding Gaussian noise, and also applied a dust extinction with $A_{\textrm{V}}=0.2$ using a \citet{Calzetti+00} extinction curve.  We blur these 2400 individual spectra to the MaNGA LSF, and then applied \textsc{Starlight} using the same SSPs and \textsc{Starlight} configuration as described in the main text to compare how well the populations are recovered under different circumstances.

\begin{figure}
    \centering
    \includegraphics[width=0.714\columnwidth]{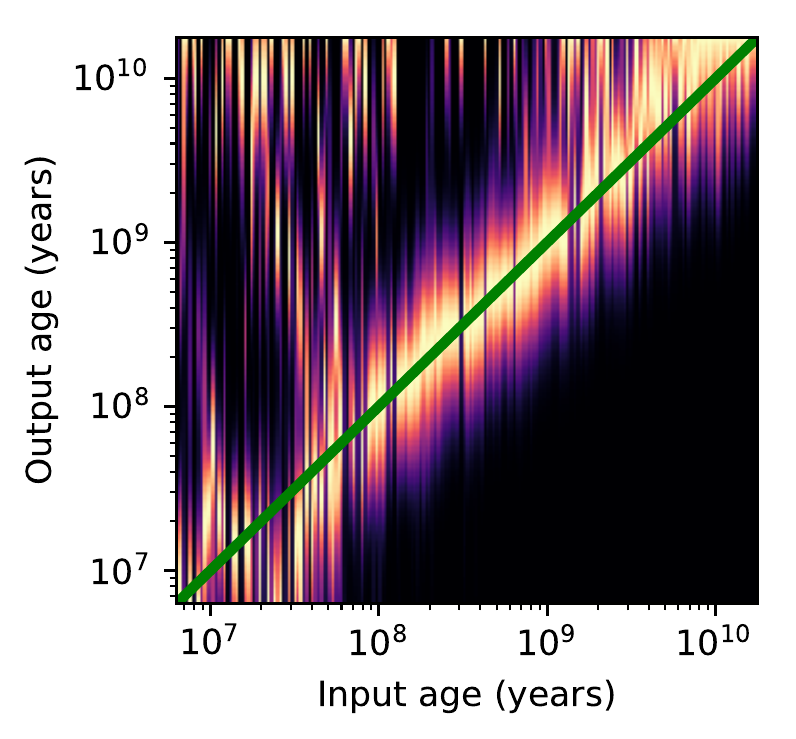}
    \caption{Distributions of the measured mass weights given by \textsc{Starlight} in fits of stellar populations of known age, for a signal-to-noise ratio of 5 and $A_{\textrm{V}}=0.2$.  Input spectra are interpolated from the grid of E-MILES SSPs at $Z = Z_{\odot}$.  The recovered weight distributions are smoothed by 0.3~dex in age, and the green line indicates equality between input and output ages.  The noise in the weights assigned to templates older than the input SSP age is due to the \textsc{Starlight} configuration used: specifically, we force 30\% of all templates to be assigned a \textit{flux} weight for consistency with the main text.  This small amount of noise in flux weights is then amplified in the conversion to mass weights.}
    \label{appfig:AgeofSSP}
\end{figure}

The 0.3-dex temporally-smoothed distributions of mass weights measured from the \textsc{Starlight} output of these known SSPs is shown in Figure~\ref{appfig:AgeofSSP} for a signal-to-noise ratios of 5 and $Z=Z_{\odot}$.  Results for a signal-to-noise ratio of 10, 15, or 20 were not noticeably improved, and providing an input metallicity of $Z=10^{-0.625}Z_{\odot}$ or $Z=10^{-1.25}Z_{\odot}$ did not change the results.  \textsc{Starlight} is able to recover the mean age of input stellar populations of all ages older than $\approx10^{8}$~years even with an input spectrum signal-to-noise ratio of 5, highlighting the diagnostic power of using such a long wavelength range to model the large-scale continuum in fitting.

A tendency for \textsc{Starlight} to also include small levels of older populations is highlighted.  This is a consequence of the combination of a strong trend in mass-to-light ratio with stellar population age and the robust \textsc{Starlight} configuration used.  As \textsc{Starlight} fits the input spectrum (i.e.\ in light space), we force it to assign a weight to at least 30\% of all SSP spectra to ensure full SFH recovery in science cases.  In the case of a single input stellar population, this will result in small spurious weights being given to other templates, and this noise becomes amplified in old populations when considering mass weights.

It is not necessarily clear how much the boundary between the two template libraries affects \textsc{Starlight}'s inability to recover stellar populations younger than $\approx 10^{7.5}$~years, or whether it is purely due to the lack of diagnostic spectral information at these ages.

\subsubsection{Effects of kinematics and dust}
\label{app:FittingTests-SSPRecovery-kindust}

To test whether these results are improved when \textsc{Starlight} does not also have to model the kinematics of the input spectrum, we repeated these test, but with the velocity and dispersion fixed to the known input values.  The results were entirely unchanged.

We also performed these same tests with $A_{\textrm{V}}=0$ and $A_{\textrm{V}}=0.8$ (instead of $A_{\textrm{V}}=0.2$ as used above) to check that \textsc{Starlight} is still able to recover populations in low- and high-extinction environments, and found the results to be unchanged here too.

\subsection{Recovery of a known star-formation history}
\label{app:FittingTests-SFHRecovery}

\begin{figure*}
    \centering
    \includegraphics[width=\textwidth]{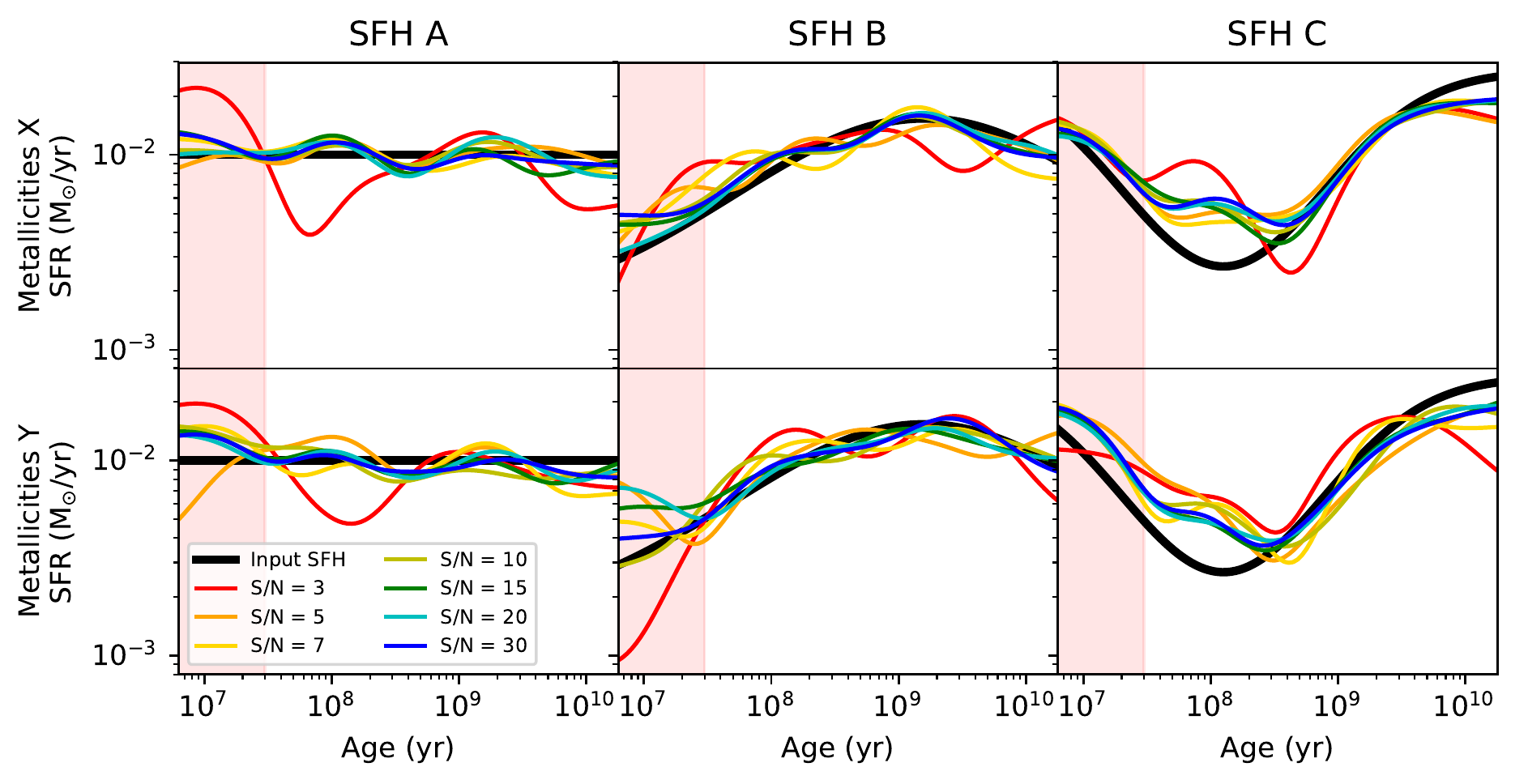}
    \caption{The measured SFHs for different input spectrum signal-to-noise ratios (coloured lines) compared to the input SFH shape (black line) for each SFH (\textit{left} to \textit{right}) and metallicity distribution (\textit{top} and \textit{bottom}).  Recovered SFHs are smoothed by 0.3~dex in age.  The red-shaded region indicates that for which SSP weights are ignored in science cases (see main text).  The general shape is recovered well in all cases, particularly for signal-to-noise ratios greater than 5.}
    \label{appfig:KnownSFH}
\end{figure*}

\begin{figure}
    \centering
    \includegraphics[width=\columnwidth]{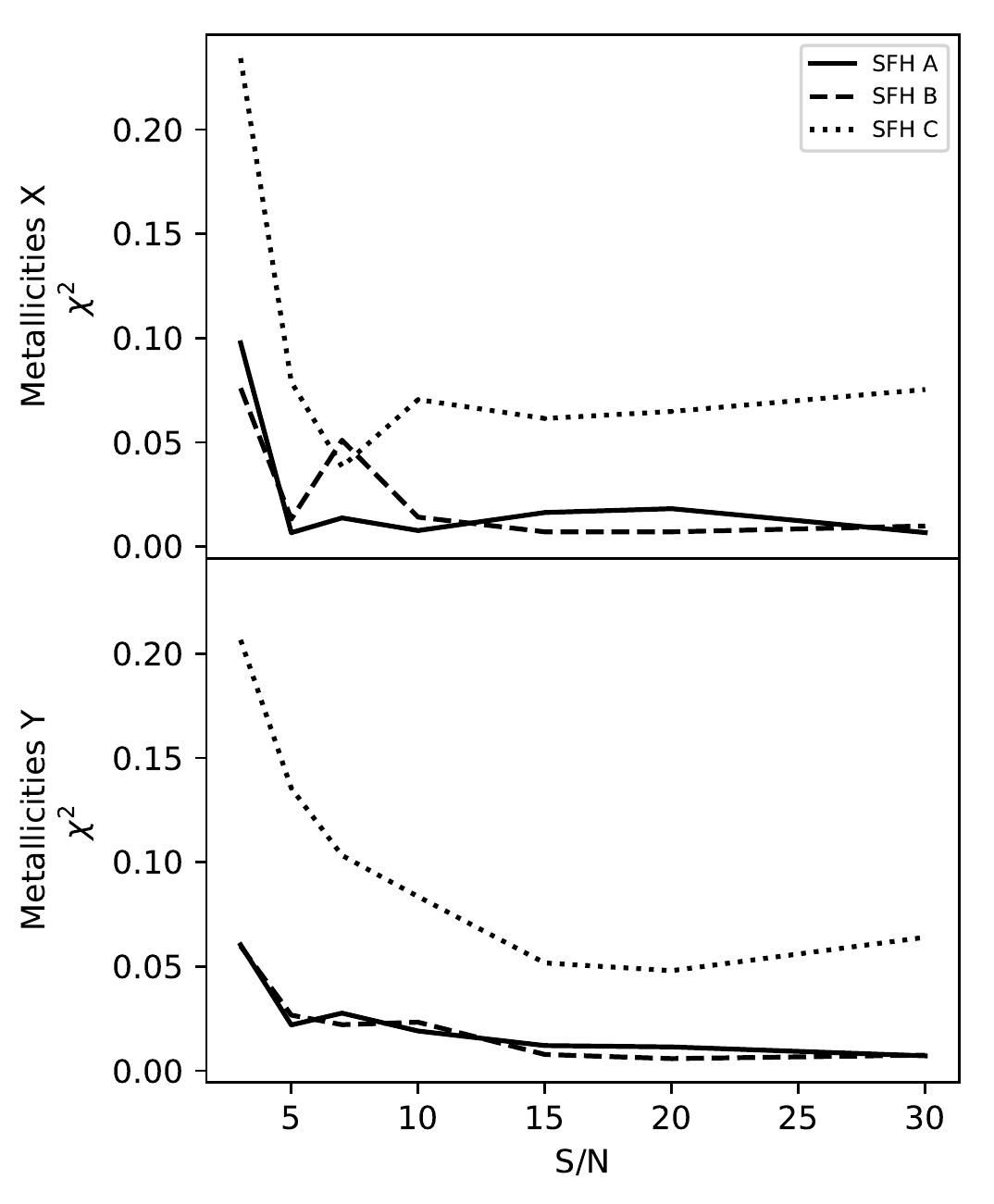}
    \caption{A $\chi^2$ goodness-of-fit measurement of the input to the measured SFH in Figure~\ref{appfig:KnownSFH}, for each of the three SFH shapes (line styles) and metallicity distributions (\textit{top} and \textit{bottom}) for different input signal-to-noise ratios.  Increasing S/N above 7 does very little to improve \textsc{Starlight}'s ability to recover the SFH.}
    \label{appfig:KnownSFH_x2}
\end{figure}

Finally, to simulate the effects of low SNR on a spectrum comprising multiple stellar populations (such as is found in real galaxies), we created spectra of three different star-formation histories (SFHs) using the template SSPs.  The different SFHs reflect different cases:
\begin{enumerate}[A:]
    \item a flat SFH, where the star-formation rate is defined as $SFR(t)~=~0.1~M_{\odot}{\rm yr}^{-1}$ for all lookback times $t$;
    \item a peaked and then declining SFH (as seen in many galaxies) with a star-formation rate represented by $SFR(t)~=~\left[0.2~+~\mathcal{N}_{9.2}^{1.5}\left(\log (t)\right)\right]~M_{\odot}{\rm yr}^{-1}$, where $\mathcal{N}_{\mu}^{\sigma}(x)$ denotes a Gaussian function of $x$ centred on $x=\mu$ with standard deviation of $\sigma$~dex;
    \item a declining and rejuvenating SFH with a star-formation history represented by $SFR(t)~=~\left[1.2-\mathcal{N}_{8.1}^{0.8}(\log (t))\right]~M_{\odot}{\rm yr}^{-1}$.
\end{enumerate}
In building these SFHs, we assign weights to each SSP assuming that they represent all star-formation between their nominal age and the next-youngest SSP.

We also included two different metallicity distributions, neither of which vary with stellar population age:
\begin{enumerate}[A:]\setcounter{enumi}{23}
    \item a flat distribution over the range of metallicities in the E-MILES templates, where the relative flux of each SSP of any given age is defined by $F_{i} = 1$;
    \item a peaked distribution where the relative flux of each SSP is defined at each age $t$ by $F_{i}~=~\mathcal{N}_{-0.71}^{0.3}(\log (Z_{\textrm{i}}/\textrm{Z}_{\odot}))$.
\end{enumerate}

These spectra were then degraded to different SNRs of $\textrm{S/N}=3$, 5, 7, 10, 15, 20, and 30.  When each of these 42 spectra of known SFHs were fit using \textsc{Starlight} in the same configuration used in the main text, we found the general shape of $SFR(t)$ is recovered in all cases, as shown in Figure~\ref{appfig:KnownSFH}.  For a SNR greater than 5 (the lowest typically found in the outskirts of MaNGA galaxies in the Primary or Primary+ samples; see \citealt{Yan+16-design}), the derived SFHs show very good agreement (with a value of $SFR(t)$ within 0.15~dex of the known value at all lookback times $t$) to the input SFHs.  This is also shown in Figure~\ref{appfig:KnownSFH_x2}, where a simple $\chi^2$ goodness-of-fit measurement is obtained between the input SFH and each of the recovered SFHs.  \textsc{Starlight} is able to recover SFHs~A and B for all cases with $\textrm{S/N}\geq10$, and while the measurement of SFH~C is worse, it is recovered equally well for $\textrm{S/N}\geq7$ in metallicity distribution~X and $\textrm{S/N}\geq15$ for metallicity distribution~Y.

\subsection{Implications}
\label{app:Implications}

We assume throughout this work that the E-MILES model spectra are accurate representations of the stellar populations they represent.  A full test of whether this is indeed the case (as conducted by e.g.\ \citealt{Ge+19}) is beyond this study, but the tests shown here imply that if this is true, we expect \textsc{Starlight} to be able to recover the true SFHs under all the conditions analysed in this work.  In fact, we find evidence that fitting spaxels individually --- rather than summing spectra from neighbouring spaxels --- may be the most robust approach to avoid the dominance of systematics from compromising the ability to measure a SFH.  Notwithstanding this robustness, to ensure that the low signal-to-noise regions of the galaxy are not affecting our results in ways we don't anticipate, in all stages of the science analysis shown in this work, we ensure that we weight spaxels by their flux or mass, ensuring that the central regions (with higher SNR and therefore with probably good fits) are emphasised, and low signal-to-noise regions are down-weighted.

% Don't change these lines
\bsp	% typesetting comment
\label{lastpage}
\end{document}